\documentclass[a4paper, 12pt, oneside]{article}

\usepackage{jheppub}
\usepackage{graphicx,wrapfig,float,slashed,subcaption,bbold,bm}
\usepackage{amsmath,amssymb,epsfig,graphicx,xcolor}
\usepackage{booktabs}
\usepackage{ragged2e}
\usepackage{mathtools}
\usepackage{cancel}
\usepackage{makecell}
\usepackage{multirow}
\usepackage{adjustbox}
\usepackage{comment}
\allowdisplaybreaks

\def\beq{\begin{equation}}
\def\eeq#1{\label{#1}\end{equation}}
\def\eeqn{\end{equation}}
\def\beqa{\begin{eqnarray}}
\def\eeqa#1{\label{#1}\end{eqnarray}}
\def\eeqan{\end{eqnarray}}

\def\leqn#1{(\ref{#1})}

\newcommand{\nc}{\newcommand}
\nc{\ttbar}{t\bar t}
\nc{\eeprocess}{e^+e^-\to t\bar t}
\def\gev{\;\mathrm{GeV}}
\def\tev{\;\mathrm{TeV}}
\def\ab{\;\mathrm{ab}^{-1}}
\def\fb{\;\mathrm{fb}^{-1}}
\newcommand{\Sec}[1]{Sec.~\ref{#1}}
\newcommand{\Fig}[1]{Figure~\ref{#1}}
\newcommand{\Eq}[1]{Eq.~(\ref{#1})}
\newcommand{\Tab}[1]{Table~\ref{#1}}

\newcommand{\tr}{\mathrm{tr}}

\newcommand{\khat}{\hat k}
\newcommand{\rhat}{\hat r}
\newcommand{\nhat}{\hat n}
\newcommand{\sig}{\sigma}
\def\lsim{\mathrel{\raise.3ex\hbox{$<$\kern-.75em\lower1ex\hbox{$\sim$}}}}
\def\gsim{\mathrel{\raise.3ex\hbox{$>$\kern-.75em\lower1ex\hbox{$\sim$}}}}

\title{Probing CP-violating top-quark dipole moments with tomographic observables}

\author[a]{Maxim Perelstein,}
\author[a,b]{Minal Shaik,}
\author[a,c]{and Taewook Youn}
\affiliation[a]{Laboratory for Elementary Particle Physics \\ Cornell University, Ithaca, NY 14853, USA}
\affiliation[b]{Abdus Salam Centre for Theoretical Physics\\ Imperial College London, London SW7 2AZ, UK}
\affiliation[c]{School of Physics, Korea Institute for Advanced Study \\ Seoul 02455, Republic of Korea}
\emailAdd{m.perelstein@cornell.edu}
\emailAdd{shaikm1@uci.edu}
\emailAdd{taewook.youn@cornell.edu}

\abstract{Quantum tomography program reconstructs the full spin density matrix of top-quark pairs from dilepton angular correlations at both electron-positron and hadron colliders. We use symmetry arguments to identify tomographic observables sensitive to CP violation in top quark couplings. As a concrete example, we study the sensitivity of these observables to (chromo-)electric dipole moment operators at the LHC and the FCC-ee. We also construct the optimal observable, which combines tomographic and kinematic information to achieve statistically optimal sensitivity for each operator and production process. The analysis is based on fast detector simulation of top pair production and decay in the dilepton channel. We project a $95\%$~CL sensitivity to new-physics scales (the inverse dipole couplings) of $10$--$15\tev$ for the electroweak electric dipoles at the FCC-ee ($365\gev$, $3\ab$) and of $40\,(200)\tev$ for the top chromo-electric dipole moment at the LHC Run~2 (HL-LHC). This sensitivities exceed those of the traditional observables currently used by experiments to use for these operators, indicating the power of the tomographic approach to search for CP violation.}

\begin{document}
\maketitle
\flushbottom

\section{Introduction}
\label{sec:intro}
\label{sec:operators}

In the Standard Model (SM), the only source of violation of CP symmetry is the Cabibbo--Kobayashi--Maskawa (CKM) phase. Extensions of the SM generically contain new sources of CP violation. New CP-violating physics around the electroweak scale is particularly well-motivated by the electroweak baryogenesis (EWB) scenario, since the SM CKM phase falls many orders of magnitude short of the CP violation required to generate the observed baryon asymmetry of the universe~\cite{Sakharov:1967dj,Farrar:1993hn}.  
The top quark is a natural place to look for these effects:
with an order-one Yukawa coupling it is the SM particle most strongly tied to
the electroweak-symmetry-breaking sector, where many baryogenesis scenarios
locate the new CP phases, and its anomalous couplings remain among the most
weakly constrained~\cite{Atwood:2000tu,Kamenik:2011dk}. The CP-violating
observables considered in this paper --- the
electric and weak dipole moments of the top and its chromo-electric dipole
moment (CEDM) --- are generated in the SM only at three loops and
beyond~\cite{Czarnecki:1997bu,Gupta:2009wu}, far
below any conceivable experimental sensitivity, so any observed signal would
be an unambiguous discovery of new physics.

The recent observation of quantum entanglement in top-quark
pairs by ATLAS and CMS~\cite{ATLAS:2023fsd,CMS:2024pts} is the culmination
of a two-decade program of $\ttbar$ spin-correlation
theory~\cite{Kane:1991bg,Bernreuther:1993hq,Mahlon:1995zn,Bernreuther:2015yna}
and measurement~\cite{ATLAS:2012ao,ATLAS:2019zrq,CMS:2019nrx}. 
It has established that the full spin density matrix of the $\ttbar$ system is experimentally accessible: the charged leptons of dileptonic decays act as perfect spin
analyzers~\cite{Czarnecki:1990pe}, and their angular correlations
reconstruct, event by event, the
two-qubit state in which the pair is produced~\cite{Afik:2020onf}. This
\emph{quantum tomography} of $\ttbar$ production --- and of two-particle
spin states more broadly~\cite{Rahaman:2021fcz} --- has so far been exploited
mainly for quantum-information questions, such as measurements of entanglement, discord and
steering, as well as tests of Bell
inequalities~\cite{Afik:2022kwm,Fabbrichesi:2021npl,Severi:2021cnj,Aguilar-Saavedra:2022uye,Afik:2022dgh,Han:2024ugl}
(see Ref.~\cite{Barr:2024djo} for a review). Searches for 
CP-\emph{even} new-physics effects in the SM effective field
theory at the LHC~\cite{Aoude:2022imd,Severi:2022qjy,Maltoni:2024tul} and
at future lepton colliders~\cite{Maltoni:2024csn,Cao:2025xnp} have also been discussed. Yet the same density matrix is
also a sharply targeted CP-violation observable: as we review below, CP
symmetry acts on the Fano coefficients of $\rho_{\ttbar}$ in a simple and
rigid way~\cite{Bernreuther:1992be,Bernreuther:1993hq}, so that CP-odd
couplings populate entries of the density matrix
that are exactly zero otherwise.

In this paper we develop this observation into a complete, quantitatively
validated program for probing the CP-violating top dipole moments at the two
colliders where they are best measured: the electroweak dipoles $d_\gamma$
and $d_Z$ in $e^+e^-\to\ttbar$ at the FCC-ee threshold run, and the
chromo-electric dipole $d_g$ in $pp\to\ttbar$ at the LHC.\footnote{The chromo-electric dipole moment of the top is also targeted by a direct CMS search~\cite{CMS:2022quh}. Other studies in recent literature related to our paper include: Ref.~\cite{Garosi:2023yxg} for the analysis of global SMEFT constraints on the top dipole operators, Ref.~\cite{Bisal:2025jwv} for a model interpretation of the CEDM, Ref.~\cite{Han:2024gan} for the study of the
electroweak dipoles at a multi-TeV muon collider, and Refs.~\cite{Janot:2015yza,Bernreuther:2017cyi} for CP-violating top couplings within the
broader top electroweak program at lepton
colliders.} Our analysis
combines three elements. First, we compute the $\ttbar$ production spin density matrix
in closed form for $e^+e^-$, $q\bar q$ and $gg$ initial states, at
the SM, dipole-SM interference, and dipole-squared orders. 
Second, from these matrix elements we construct the statistically optimal CP-odd observable. We then determine the sensitivity of this observable within an idealized parton-level analysis, and compare it to observables given by simple tomographic projections.  
Third, we carry the
entire program to detector level with measured reconstruction efficiencies, per-observable dilutions and selection optimizations at both colliders. Along the way we prove a no-go result for a
recently proposed quantum-information CP probe: the top--antitop discord
asymmetry~\cite{Afik:2022dgh,Subba:2026nzs} identically vanishes for the dipole couplings, and therefore cannot be used to probe this form of CP violation.

We parametrize the anomalous top-quark dipole couplings to the electroweak gauge
bosons ($V=\gamma,Z$) and to the gluon by two dipole Lagrangians,
\begin{align}
\label{eq:Lew}
\mathcal{L}_{\rm EW} &= -\frac{1}{\sqrt2}\sum_{V=\gamma,Z}
   \bar t\,\sigma^{\mu\nu}\big(a_V^{}+ i\,d_V^{}\,\gamma_5\big)\,t\;F^{V}_{\mu\nu}, \\[3pt]
\label{eq:Lqcd}
\mathcal{L}_{\rm QCD} &= -\frac{1}{\sqrt2}\,g_s\,
   \bar t\,\sigma^{\mu\nu}T^a\big(a_g^{}+ i\,d_g^{}\,\gamma_5\big)\,t\;G^{a}_{\mu\nu},
\end{align}
with $\sigma^{\mu\nu}=\tfrac{i}{2}[\gamma^\mu,\gamma^\nu]$, $F^V_{\mu\nu}$ the
$\gamma/Z$ field strengths, $G^a_{\mu\nu}$ the gluon field strength, $T^a$ the
color generators and $g_s$ the strong coupling. The couplings
$a_V,d_V,a_g,d_g$ are \emph{dimensionful}, of mass dimension $-1$, so each inverse
coupling $1/d$ is an energy scale; we quote all sensitivities directly as the
$95\%$ CL reach on $1/d$ in TeV. For each gauge boson the Hermitian ($\propto a$) term is
the CP-conserving \emph{magnetic} (chromo-magnetic) dipole, and the $\gamma_5$
($\propto d$) term the CP-violating \emph{electric} (chromo-electric) dipole; $d_g$ is the
top chromo-EDM. The electroweak dipoles $a_V,d_V$ are probed in
$e^+e^-\to\ttbar$ (\Sec{sec:ee}) and the chromo-dipole $a_g,d_g$ in $pp\to\ttbar$ (\Sec{sec:lhc}). Since we are interested in CP violation, we focus throughout on the electric
dipoles $d_\gamma,d_Z,d_g$; the CP-conserving magnetic couplings only rescale the
symmetric spin correlations and the rate --- e.g.\ the entanglement marker
$D=-\tfrac13\tr C$ of Ref.~\cite{Afik:2020onf} --- and belong to the entanglement
program (\Sec{sec:qinfo}). 
In our Monte Carlo (MC) study, we implement \Eq{eq:Lew} and \Eq{eq:Lqcd} in
\texttt{FeynRules}~\cite{Alloul:2013bka}, export a \texttt{UFO}
model~\cite{Degrande:2011ua}, and generate events with
\texttt{MadGraph5\_aMC@NLO}~\cite{Alwall:2014hca}.

The paper is organized as follows. Section~\ref{sec:framework} develops the
CP and quantum-information structure of the $\ttbar$ spin density matrix and
defines the CP-odd observables, including the optimal observable constructed
from the Fano coefficients. Section~\ref{sec:reco} discusses how the
observables are reconstructed from the top decay products, and
Section~\ref{sec:sim} describes the simulation framework and statistical procedure.
Sections~\ref{sec:ee} and~\ref{sec:lhc} present the analyses at the $e^+e^-$
collider and the LHC, respectively, from parton-level analyzing power to
detector-level reach. We conclude in Section~\ref{sec:conclusions}. The
derivation and optimality proof of the optimal observable are collected in
Appendix~\ref{app:oopt}, and the exact analytic form of the spin density
matrix for all three production channels is given in Appendix~\ref{app:rho}.

\section{CP violation and quantum observables}
\label{sec:framework}

\subsection{Fano decomposition of the $\ttbar$ spin state}
\label{sec:fano}

\begin{figure}[t]
\centering
\includegraphics[width=0.72\textwidth]{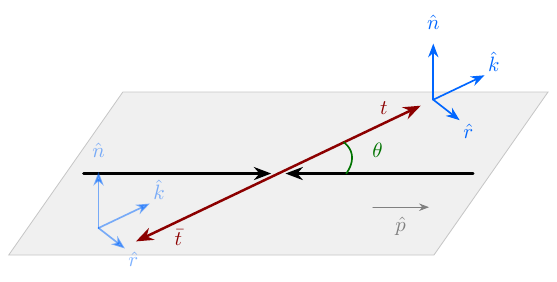}
\caption{The helicity basis of \Eq{eq:krn} in the $\ttbar$
center-of-mass frame: $\khat$ is the top direction, $\rhat$ completes the
basis within the production plane (shaded), and $\nhat$ is the $\tilde T$-odd
normal to the plane; $\theta$ is the angle between $\khat$ and the reference
axis $\hat p$ of \Eq{eq:krn}. A single triad is defined per event, and both
the top and antitop spins are projected on it.}
\label{fig:axes}
\end{figure}

The $\ttbar$ pair is a two-qubit system. Its production spin density matrix
admits the Fano decomposition
\begin{equation}
\label{eq:rho}
\rho=\frac14\Big[\mathbb{1}\otimes\mathbb{1}
 +\sum_i B_i^{+}\,\sigma_i\otimes\mathbb{1}
 +\sum_j B_j^{-}\,\mathbb{1}\otimes\sigma_j
 +\sum_{ij} C_{ij}\,\sigma_i\otimes\sigma_j\Big],
\end{equation}
with $B_i^{+}=\langle\sigma_i\otimes\mathbb1\rangle$ the top polarization,
$B_j^{-}$ the antitop polarization, and $C_{ij}=\langle\sigma_i\otimes\sigma_j
\rangle$ the spin correlation matrix. We work in the helicity basis $\{\khat,\rhat,\nhat\}$, shown in Fig.~\ref{fig:axes}:
\begin{equation}
\label{eq:krn}
\khat=\hat p_t,\qquad
\rhat=\frac{\hat p-\cos\theta\,\khat}{\sin\theta},\qquad
\nhat=\frac{\hat p\times\khat}{\sin\theta},\qquad
\cos\theta=\khat\cdot\hat p\,.
\end{equation}
Here $\hat{p}_t$ is the direction of the top quark in the 
$\ttbar$ center-of-mass (CM) frame; $\hat{p}$ is the reference axis --- the electron direction in
$e^+e^-\to\ttbar$ and the beam axis in $pp\to\ttbar$ (with the
$\mathrm{sgn}(\cos\theta)$ convention of \Sec{sec:lhc} for the $z\to-z$ symmetric $pp$ initial state); and $\nhat$ is normal to the production plane (the T-odd direction). 
Throughout, $\theta$ is the production angle and $\beta$ the top velocity in the $\ttbar$ center-of-mass frame. A single $\{\khat,\rhat,\nhat\}$ triad is built from
each event's top direction, and the spins of \emph{both} the top and the antitop are projected onto it. The explicit expressions for the Fano coefficients $B$ and $C$ in the SM, and the corrections due to the dipole moment operators,
are collected in Appendix~\ref{app:rho}.

\subsection{CP structure}
\label{sec:cpstructure}

CP-violating interactions populate the spin density matrix in a certain pattern, distinct from that of CP-conserving physics. This pattern can be understood from symmetry considerations. In the
$\ttbar$ center-of-mass frame, CP interchanges the top and antitop. With the
antitop-conjugated analyzer convention of \Sec{sec:decaymethod} below, this maps the
Fano coefficients as 
\beq
\vec B^+\leftrightarrow\vec B^-~~~~~{\rm and}~~~~~
C\leftrightarrow C^{T}, 
\eeq{FanoCP}
so CP invariance forces $\vec B^+=\vec B^-$ and a \emph{symmetric} correlation matrix. 
A CP-odd coupling therefore populates, at
linear order, only the CP-odd combinations 
\beq
\Delta B_i=B_i^+-B_i^-~~~~{\rm and}~~~~~A_{ij}=C_{ij}-C_{ji}. 
\eeq{FanoCPodd}
These combinations identically vanish in the SM (up to completely negligible three-loop corrections) as well as in any CP-conserving extension thereof, making them uniquely suited to search for CP violation. 

Naive time reversal $\tilde T$ (reversal of momenta
and spins without exchange of initial and final states) refines the pattern
further. Among the axes $\{\khat,\rhat,\nhat\}$ only
$\nhat\propto\hat p\times\khat$ is $\tilde T$-odd, so at tree level, where
the amplitudes carry no absorptive phases, the CP-odd observables generated
by the \emph{real} (Hermitian) electric dipoles are 
\beq
\Delta B_n,~~~~~A_{rn},~~~~~{\rm and}~~~~~A_{kn}. 
\eeq{FanoCPtildeTodd}
The $\tilde T$-even CP-odd combinations ($\Delta B_{k,r}$, $A_{kr}$) require
absorptive parts or anti-Hermitian couplings and vanish at tree
level~\cite{Bernreuther:1993hq}.

We have verified explicitly that the symmetry-based pattern described above is indeed realized within the specific model of new physics studied here, Eqs.~\leqn{eq:Lew} and~\leqn{eq:Lqcd}. This was confirmed both analytically (from the Fano coefficients listed in Appendix~\ref{app:rho}) and numerically, using the Monte Carlo implementation of teh model describeb below. 
The (chromo-)magnetic dipole interactions $a_\gamma,a_Z,a_g$ rescale the symmetric part of $C$ relative to the SM prediction, and generate $B_{k}$ and $B_{r}$ with $B^+=B^-$.
The CP-violating (chromo-)electric dipoles $d_\gamma,d_Z,d_g$ are the sole source of the antisymmetric
normal-plane entries $C_{kn}=-C_{nk}$ and $C_{rn}=-C_{nr}$, as well as\footnote{Non-vanishing normal polarizations are only generated at $e^+e^-$. They require interference with an axial coupling, which the pure-vector gluon
vertex cannot supply.} $\Delta B_n$. The observables in Eq.~\leqn{FanoCPtildeTodd} thus provide a clean and unambiguous probe of CP violation, as mandated by the symmetry arguments.

\subsection{Quantum-information observables: entanglement and discord}
\label{sec:qinfo}

The production spin density matrix $\rho$ is the
object from which the quantum-information content of the $\ttbar$ pair is
extracted. \emph{Entanglement} is diagnosed by the concurrence $\mathcal{C}(\rho)$
or, near threshold, by the marker $\Delta=-(C_{kk}+C_{rr}+C_{nn})-1>0$; the
$\ttbar$ state produced just above threshold is close to the maximally entangled
singlet and is the target of the ATLAS and CMS entanglement
measurements~\cite{ATLAS:2023fsd,CMS:2024pts}.

\emph{Quantum discord} quantifies the quantum correlations that persist even in
separable states~\cite{Ollivier:2001fdq,Henderson:2001wrr}; for $\ttbar$ it
was studied in Refs.~\cite{Afik:2022dgh,Han:2024ugl}. The total correlation
in $\rho$ is the quantum mutual information
\begin{equation}
\label{eq:mutinfo}
\mathcal{I}(\rho)=S(\rho_t)+S(\rho_{\bar t})-S(\rho),
\qquad S(\rho)=-\tr\rho\log_2\rho ,
\end{equation}
with $\rho_t=\tr_{\bar t}\rho$, $\rho_{\bar t}=\tr_t\rho$ the reduced states. The
part of $\mathcal{I}$ that is locally accessible by a complete projective
measurement $\{\Pi_k^{\bar t}\}$ on the antitop defines the classical correlation
\begin{equation}
\label{eq:classcorr}
\mathcal{J}_{\bar t}(\rho)=S(\rho_t)-\min_{\{\Pi_k^{\bar t}\}}\sum_k p_k\,S(\rho_t^{\,k}),
\qquad
p_k=\tr(\Pi_k^{\bar t}\rho),\quad
\rho_t^{\,k}=\frac{\tr_{\bar t}(\Pi_k^{\bar t}\rho\,\Pi_k^{\bar t})}{p_k},
\end{equation}
and the (one-way) quantum discord obtained by measuring the antitop is the
difference between the total and classical correlations,
\begin{equation}
\label{eq:discord}
\mathcal{D}(t)\equiv \mathcal{I}(\rho)-\mathcal{J}_{\bar t}(\rho)
=S(\rho_{\bar t})-S(\rho)+\min_{\{\Pi_k^{\bar t}\}}\sum_k p_k\,S(\rho_t^{\,k}) .
\end{equation}
It measures the correlation that cannot be extracted by any local measurement, and
a closed-form expression exists for two qubits~\cite{Luo:2008ecu}. $\mathcal{D}(\bar t)$ is
defined symmetrically by instead measuring the top. Discord is nonzero for a far
larger class of states than entanglement, and probing it in $\ttbar$ has been
proposed as a route to detect genuinely quantum correlations beyond
entanglement~\cite{Han:2024ugl}. All of $\mathcal{C}$, $\mathcal{D}(t)$ and
$\mathcal{D}(\bar t)$ are functions of the Fano coefficients $B^\pm$ and $C$ of \Eq{eq:rho}, and so
come ``for free'' once the tomography of \Eq{eq:BC} is done.

\paragraph{Discord does not probe CP violation in the top dipoles.}
Because measuring the antitop is not the same as measuring the top, the discord is
in general asymmetric, $\mathcal{D}(t)\neq\mathcal{D}(\bar t)$. Since CP exchanges
the two subsystems, the asymmetry
\begin{equation}
\label{eq:deltaD}
\Delta\mathcal{D}\equiv \mathcal{D}(t)-\mathcal{D}(\bar t)
\end{equation}
is CP-odd by construction and vanishes when CP is conserved; it was
introduced for $\ttbar$ in Ref.~\cite{Afik:2022dgh} and recently advocated
as ``a direct and theoretically transparent probe of
CP-violating interactions''~\cite{Subba:2026nzs}. We find, however, that
$\Delta\mathcal{D}$ is \emph{identically zero} for the top electroweak dipoles in both
$\eeprocess$ and $pp\to t\bar{t}$. Computing $\mathcal{D}(t)$ and $\mathcal{D}(\bar t)$ from the
analytic $\rho(\theta,\beta)$ --- optimizing the conditional entropy over all
local projective measurements --- and scanning $\theta$, $\beta$, and each coupling,
we find $\Delta\mathcal{D}$ consistent with zero to machine precision, against an
exactly vanishing SM baseline, including for the CP-violating electric couplings
$d_Z,d_\gamma$. The dipoles do shift the individual discords (and the
concurrence) at the few-percent level, but $\mathcal{D}(t)$ and
$\mathcal{D}(\bar t)$ move together, so their difference carries no signal. This is
not a numerical accident, but rather a structural consequence of the form of the spin-density matrix under consideration:

\begin{quote}
\emph{Proposition.} If (i) the top and antitop acquire equal-magnitude,
opposite-sign normal polarizations, $B^{+}_n=-B^{-}_n$, with equal $k,r$
components ($B^{+}_{k,r}=B^{-}_{k,r}$); and (ii) the antisymmetric part of $C$ is
confined to the normal plane, $C_{kn}=-C_{nk}$, $C_{rn}=-C_{nr}$, with the $k$--$r$
block symmetric; then $\mathcal{D}(t)=\mathcal{D}(\bar t)$ identically, i.e.\
$\Delta\mathcal{D}=0$.
\end{quote}

\noindent\emph{Proof.} Conditions (i)--(ii) make $\rho$ invariant under the
combined operation $G=\mathcal{S}\,(K_t\otimes K_{\bar t})$, where $\mathcal{S}$
swaps the two qubits and $K$ is the single-qubit antiunitary (complex conjugation
in the frame in which the T-odd normal axis $\nhat$ is the imaginary direction)
that reflects $\nhat\to-\nhat$ while leaving $\khat,\rhat$ invariant. Under $\mathcal{S}$ the
polarizations exchange and $C\to C^{T}$; under $K_t\otimes K_{\bar t}$ the
$\nhat$-odd entries ($B_n^\pm$ and $C_{kn},C_{nk},C_{rn},C_{nr}$) flip sign.
Applying $G$ and using (i)--(ii) returns $\rho$ term by term, so $G(\rho)=\rho$.
Now $\mathcal{S}$ maps $\mathcal{D}(t)$ into $\mathcal{D}(\bar t)$, while
$K_t\otimes K_{\bar t}$ is a \emph{local} antiunitary and therefore preserves every
von Neumann entropy in \Eq{eq:discord}, and hence the discord itself. Thus
$\mathcal{D}(t)=\mathcal{D}(\bar t)$. $\qquad\blacksquare$

Both conditions of the above Proposition hold for the Fano coefficients in our analysis. In fact, they hold term by term --- in the SM, the linear (SM-dipole interference) piece,
and the dipole-squared piece --- even if both CP-conserving and CP-violating dipole operators are turned on simultaneously. In fact, addition of {\it any} Hermitian operator to the SM does not violate these conditions at tree level.\footnote{Retaining the absorptive part of the $Z$ propagator, which we neglected, breaks the
conditions of the Proposition: the SM then acquires a normal polarization $B^{+}_n=B^{-}_n\neq0$ and the dipole generates in-plane components of $\vec B^{+}-\vec B^{-}$. Both are suppressed by $\Gamma_Z m_Z/(s-m_Z^2)\sim2\times10^{-3}$, leaving $\Delta\mathcal{D}$ four to five orders of magnitude below $A_{rn}$, far beyond any conceivable experimental sensitivity.} Hence $\Delta\mathcal{D}$ receives \emph{no} contribution from new physics, and cannot serve as a probe of CP violation.

The validity of the above no-go theorem is not limited to the entropic discord of~\Eq{eq:discord}: the proof uses
only that $\mathcal{D}$ is invariant under local (anti)unitaries and is exchanged
with its $t\leftrightarrow\bar t$ partner by the swap $\mathcal{S}$. The same two
properties hold for the \emph{geometric} quantum discord --- the Hilbert--Schmidt
distance from $\rho$ to the nearest classical--quantum state~\cite{Dakic:2010xfz},
which is the measure actually used in Ref.~\cite{Subba:2026nzs} --- since the
Hilbert--Schmidt norm is preserved by (anti)unitaries and the set of
classical--quantum states is swap-covariant. Thus the geometric discord asymmetry
$\Delta\mathcal{D}_G$ vanishes identically as well, and our conclusion is
independent of which discord measure is adopted.

Indeed, invariance under local (anti)unitaries and swap-covariance are shared by
\emph{every} standard measure of bipartite correlations, so the same $G$-invariance
of $\rho$ annihilates the corresponding $t\leftrightarrow\bar t$ asymmetry. A
natural candidate, given that steering is intrinsically one-way, is the
EPR-steering~\cite{Wiseman:2007hyt} asymmetry
$\Delta\mathcal{S}\equiv\mathcal{S}(t\!\to\!\bar t)-\mathcal{S}(\bar t\!\to\! t)$;
it too vanishes identically for the dipoles.\footnote{Explicitly, for the
steering-ellipsoid volume~\cite{Jevtic:2015epl}, $V(a\!\to\! b)\propto
|\det(C-a\,b^{T})|/(1-|a|^2)^2$, one finds $V(t\!\to\!\bar t)/V(\bar t\!\to\! t)=
[(1-|B^{-}|^2)/(1-|B^{+}|^2)]^2$, since $\det(C^{T}-B^{-}B^{+T})=\det(C-B^{+}B^{-T})$;
condition (i) gives $|B^{+}|=|B^{-}|$, so the ratio is unity. A numerical scan
confirms $\Delta\mathcal{S}=0$ to machine precision.} Entanglement, discord, and
steering are therefore all blind to the CP violation at tree level.

The above no-go theorem contradicts the central claim of Ref.~\cite{Subba:2026nzs}. In fact, the $\Delta\mathcal{D}$
distributions reported in~\cite{Subba:2026nzs} are symmetric under the sign flip of the genuinely CP-odd coupling ($C_2^{A}$), i.e.\ there is no linear response, as required by the symmetry arguments. The reported small nonzero values, $\mathcal{O}(10^{-4})$--$\mathcal{O}(10^{-6})$, are generated by CP-\emph{even} couplings, and can arise only from absorptive effects, such as the absorptive part of the $Z$ propagator. In any case, a measurement of discord to the required precision is not realistic given the experimental uncertainties. Instead, we propose that a clean, direct and robust search for CP-violating dipole interactions can be performed using the observables in Eq.~\leqn{FanoCPtildeTodd}, or their extension to the ``optimal" observable discussed in Section~\ref{sec:optimal} below. Like discord, our observables are contained within the $t\bar{t}$ spin-density matrix, and form a subset of the ``quantum tomography" program. Unlike discord, they do not have an obvious quantum-information interpretation, focusing instead on the features directly attributable to the violation of CP.

\section{Top Reconstruction and the Optimal CP-Odd Observable}
\label{sec:reco}
\label{sec:decaymethod}

Top-pair quantum tomography uses the final states where both top and antitop decay semi-leptonically. In the narrow-width approximation, production factorizes from decay, and the charged lepton of each (anti)top acts as a maximal-analyzing-power spin analyzer. We write
$\hat q_+$ for the unit direction of the $\ell^+$ in the \emph{top} rest frame
(reached from the $\ttbar$ center-of-mass frame by boosting along the top
momentum), and $\hat q_-$ for \emph{minus} the unit direction of the $\ell^-$
in the \emph{antitop} rest frame. With this notation, CP
acts simply as $\hat q_+\leftrightarrow\hat q_-$. The standard
spin-analyzer
relations~\cite{Mahlon:2010gw,Bernreuther:2015yna} then give
the Fano coefficients as
\begin{equation}
\label{eq:BC}
B_i^{+}=\frac{3}{\alpha}\langle \hat q_+^{\,i}\rangle,\qquad
B_j^{-}=\frac{3}{\alpha}\langle \hat q_-^{\,j}\rangle,\qquad
C_{ij}=\frac{9}{\alpha^2}\langle \hat q_+^{\,i}\,\hat q_-^{\,j}\rangle,
\end{equation}
where the factors $3,9$ follow from the second moments of the distribution (the
angular average $\langle\hat q_i\hat q_j\rangle=\tfrac13\delta_{ij}$). Here $i, j$ denote the components in the helicity basis, Eq.~\leqn{eq:krn}.  

The brackets in \Eq{eq:BC} are \emph{ensemble} averages: 
the density matrix is reconstructed statistically over many events,
not event by event, as is standard in $\ttbar$ spin-density-matrix
studies~\cite{Afik:2020onf,Aoude:2022imd,Severi:2022qjy,ATLAS:2023fsd,CMS:2024pts,Han:2024ugl,Cao:2025xnp}. The Fano coefficients depend on top production kinematic variables, $\theta$ and $\beta$ (the latter is fixed in the case of $e^+e^-$ production). For example, in Figures~\ref{fig:eetheta} and~\ref{fig:lhctheta} below, the CP-odd observables are plotted as a function of the top production angle in the $t\bar{t}$ rest frame. To fully reconstruct this dependence, the events can be binned in $\theta$ and $\beta$, and the averages in Eq.~\leqn{eq:BC} are computed separately in each bin~\cite{Maltoni:2024csn}. An alternative is to average over the entire sample. This yields the inclusive, rate-weighted density matrix
\begin{equation}
\label{eq:rhoincl}
\rho_{\rm incl}=\int\frac{d\sigma(\theta,\beta)}{\sigma}\,\rho(\theta,\beta),
\end{equation}
where $\sigma$ is the total cross section and $d\sigma/\sigma$ the normalized
differential rate. Since the helicity basis $\{\khat,\rhat,\nhat\}$ is rebuilt in each
event from that event's top direction, the projections co-rotate with the
kinematics and are directly comparable across events despite their differing
$(\theta,\beta)$. 

In the analyses below, we will use the inclusive, rate-weighted version of the CP-odd observables 
$A_{rn}$, $A_{kn}$ and $\Delta B_n$ defined in Eq.~\leqn{FanoCPtildeTodd}. Each inclusive observable is the sample mean of a \emph{per-event estimator} built from the angular moments of \Eq{eq:BC}: 
\begin{equation}
\label{eq:estimators}
A_{rn}=9\big\langle\hat q_+^{\,r}\hat q_-^{\,n}-\hat q_+^{\,n}\hat q_-^{\,r}\big\rangle,
\qquad
\Delta B_n=3\big\langle\hat q_+^{\,n}-\hat q_-^{\,n}\big\rangle,
\end{equation}
and similarly for $A_{kn}$. We have verified that keeping the additional information contained in the dependence of these observables on top production kinematics does not yield a significant increase in the sensitivity of the analysis.

\paragraph{Top reconstruction.} The above discussion assumed that the top and anti-top rest frames are known on an event-by-event basis. This requires determination of the four-momenta of each of the two neutrinos present in the final state, which cannot be measured directly. Instead, the momenta are inferred by a kinematic fit. The fit in the case of $e^+e^-$ collision exploits the known four-momentum of the $e^+e^-$ system,
$P_{\rm cm}=(\sqrt s,\vec 0)$. The missing four-momentum is fixed to
$p_{\nu_1}+p_{\nu_2}=P_{\rm cm}-(p_{\ell^+}+p_{\ell^-}+p_{b_1}+p_{b_2})$, and the neutrino split
is determined by minimizing
\begin{equation}
\label{eq:kinfit}
\chi^2=\sum_{W^\pm}\left(\frac{m^2_{\ell\nu}-M_W^2}{2M_W\Gamma_W}\right)^2
+\sum_{t,\bar t}\left(\frac{m^2_{b\ell\nu}-m_t^2}{2m_t\Gamma_t}\right)^2
+\sum_{i=1,2}\left(\frac{s_i-1}{\sigma_{\rm jes}}\right)^2
\end{equation}
over the neutrino momentum $\vec\nu_1$ and two $b$-jet energy-scale factors
$s_i$ (resolution $\sigma_{\rm jes}=5\%$), with $\nu_2$ fixed by momentum
conservation and a mild penalty tying the fitted neutrino energies to the
measured missing energy. Both lepton--$b$ pairings are fit, and the one that yields lower
$\chi^2$ retained. The top four-momenta are then assembled from the best-fit neutrino
momenta, $p_t=p_{b_1}(s_1)+p_{\ell^+}+p_{\nu_1}$ and likewise
for $p_{\bar t}$, and all rest-frame quantities below are computed from them. The $\chi^2$ value itself is kept only as a quality score. The widths act as weights rather than calibrated
resolutions, so $\chi^2$ is a per-event reconstruction-quality \emph{score}:
cut values on it are treated as empirical working points (\Tab{tab:cuts}), not
probabilistically. A similar procedure is followed at the LHC. However, the hadron collider fit has fewer inputs: Only the transverse components of the total missing momentum can be inferred , and the rest frame of the $t\bar{t}$ system is not known {\it a priori}. To accommodate this, the jet energy scales are not included as fit parameters at the LHC, and jet energies are instead smeared within resolution. This is the standard procedure in LHC dilepton-top reconstruction.

\subsection{Optimal CP-odd observable}
\label{sec:optimal}

Each of the three observables of Eq.~\leqn{FanoCPtildeTodd} provides a clean probe for CP violation. In this section, we will construct a function which combines the three observables, along with kinematic information about the tops and their decay products, to improve the sensitivity to the CP-odd dipole operators. In fact, our combined observable is optimal, {\it i.e.} it provides the maximal possible statistical sensitivity at the parton level, as will be proven in Appendix~\ref{app:oopt}.     

\begin{figure}[t!]
\centering
\includegraphics[width=\textwidth]{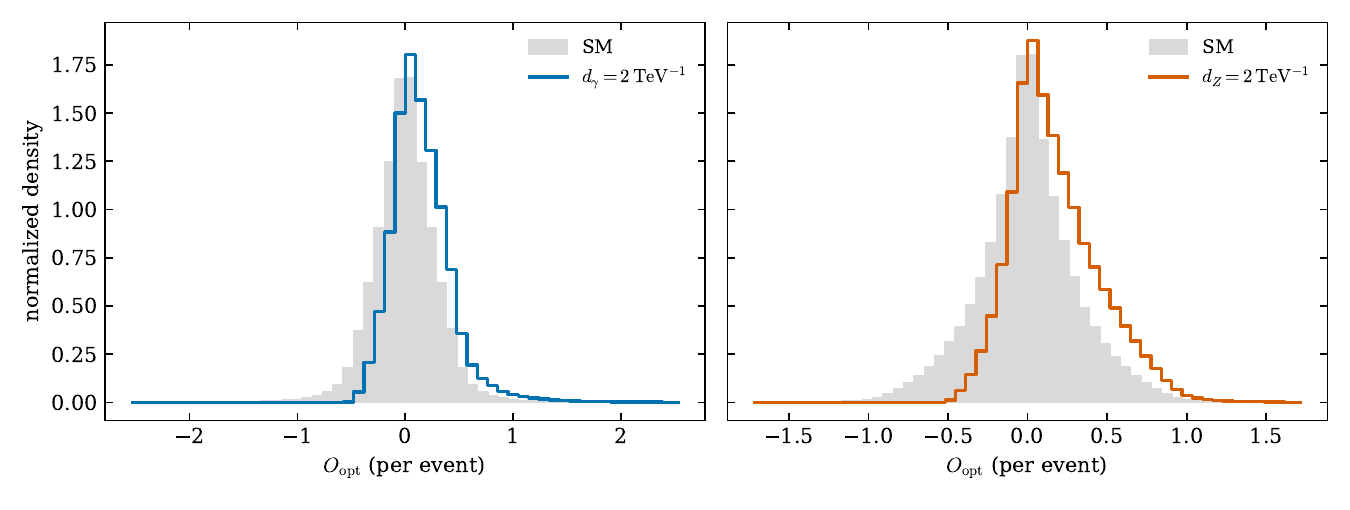}
\caption{Distribution of the per-event optimal observable
\Eq{eq:ooptfano} in $e^+e^-\to t\bar{t}$, at $\sqrt s=365\gev$, in the SM and with each electric dipole operator switched on separately, $d_\gamma$ (left) and $d_Z$ (right). The illustrative value $d_{\gamma,Z}=2\,{\rm TeV}^{-1}$
is chosen to make the effect of the dipole visible by eye.}
\label{fig:ooptdist}
\end{figure}

Writing the dilepton distribution at linear order in a dipole coupling $d$
as~\cite{Atwood:1991ka,Diehl:1993br,Bernreuther:1995nw,Bartl:1998nn,Durieux:2018tev}
\begin{equation}
\label{eq:oopt}
d\sigma=\big[f_0(\Omega)+d\,f_1(\Omega)+\mathcal O(d^2)\big]\,d\Omega,
\qquad
O_{\rm opt}(\Omega)=\frac{f_1(\Omega)}{f_0(\Omega)},
\qquad \Omega=(\cos\theta,\hat q_+,\hat q_-),
\end{equation}
the mean of the per-event weight $O_{\rm opt}$ is the statistically optimal
estimator of $d$: it saturates the Cram\'er--Rao bound, and the significance
of \emph{any} observable $O$ is ${\rm corr}_0(O,O_{\rm opt})$ times the
optimal one, the subscript denoting moments taken in the SM
(proof in Appendix~\ref{app:oopt}). Although the full density is
quadratic in the coupling, no linearization is being assumed here: $O_{\rm
opt}$ is the statistical \emph{score} at $d=0$, the locally most powerful
test of the SM hypothesis, so its definition involves only the derivative
$f_1$ by construction. At the sensitivity frontier the neglected quadratic term is in any case negligible.\footnote{Denoting the $\mathcal O(d^2)$ density coefficient as $f_2$ and  the $95\%$~CL bound of~\Sec{sec:sim} as $d_{95}$, we obtain from numerical simulation at FCC-ee that $d^2f_2/f_0\sim2\times10^{-4}$ at $d=d_{95}$, against $d\,f_1/f_0\sim3\times10^{-2}$ for the signal itself. The estimates for the LHC are of the same order.} It is important that
$d\sigma$ is \emph{not} measured event by event: the densities $f_{0,1}$ are
known theoretical functions --- the Fano coefficients of \Eq{eq:rho} in the SM
and at linear order in the dipole, contracted with the spin analyzers --- and
the event supplies only the point $\Omega$ at which to evaluate them.
Explicitly,
\begin{equation}
\label{eq:ooptfano}
O_{\rm opt}(\Omega)=
\frac{\vec B^+_1\!\cdot\hat q_+ + \vec B^-_1\!\cdot\hat q_-
      + \hat q_+\!\cdot C_1\,\hat q_-}
     {1+\vec B^+_0\!\cdot\hat q_+ + \vec B^-_0\!\cdot\hat q_-
      + \hat q_+\!\cdot C_0\,\hat q_-}\,,
\end{equation}
where the subscripts $0,1$ denote the SM and linear-dipole Fano coefficients
--- known analytic functions of $\cos\theta$ (and of $m_{t\bar t}$ at a hadron
collider) --- while $\hat q_\pm$ are measured for each event. For a CP-odd dipole, $C_1$ and $B^\pm_1$ have the structure discussed in Section~\ref{sec:cpstructure}, and $O_{\rm opt}$ 
becomes a linear combination of the observables
of~\Eq{FanoCPtildeTodd}, combining them with their optimal angle-dependent weights. As an example, the distributions of $O_{\rm opt}$ on a Monte Carlo sample of $e^+e^-\to t\bar{t}$ events at $\sqrt{s}=365$~GeV, in the SM and in the presence of a CP-violating dipole operator, are shown in Figure~\ref{fig:ooptdist}. The SM distribution is symmetric around zero, and has vanishing mean, while 
the CP-odd dipole skews the distribution, generating a non-zero mean proportional to the strength of the dipole. Further details, including the correlation coefficients between $O_{\rm opt}$ and its input observables, are provided in Appendix~\ref{app:oopt}.

\section{Simulation framework and statistical procedure}
\label{sec:sim}

All Monte-Carlo samples used below are generated with
\texttt{MadGraph5\_aMC@NLO}~\cite{Alwall:2014hca} from the \texttt{UFO} models incorporating the dipole operators
of Eqs.~\leqn{eq:Lew},~\leqn{eq:Lqcd}. We simulate $t\bar{t}$ production at the LHC ($pp$, $\sqrt{s}=13$~TeV) and a future electron-positron collider ($e^+e^-$, $\sqrt{s}=365$~GeV$-1$~TeV), followed by the dilepton decay of the $t\bar{t}$ pair. We perform both parton-level and parametrized detector-level analyses. For the latter, the events are showered with \texttt{Pythia8}~\cite{Sjostrand:2014zea}, and passed through the \texttt{Delphes}~\cite{deFavereau:2013fsa} fast detector simulation --- the
\texttt{IDEA} card at FCC-ee and the \texttt{CMS} card at the LHC. 

The number of events available in each analysis can be written as
\beq
N=\varepsilon\,\sig_{\rm dilep}\mathcal L\,,
\eeq{Nfla}
where $\mathcal L$ is the integrated luminosity, $\sig_{\rm dilep}$
is the $t\bar{t}$ production cross section multiplied by the branching ratio of the dilepton decay channel, and the reconstruction--selection efficiency $\varepsilon$ is the fraction of dilepton events in which the detector reconstructs both leptons and the required ($b$-)jets, the selection cuts are passed, and the kinematic fit (\Sec{sec:detector}, \Sec{sec:lhc}) returns a physical solution. For a parton-level analysis with no selection cuts, $\varepsilon=1.0$. 

Consider a CP-odd observable $\mathcal O$, which can be either one of the observables defined in~\Eq{FanoCPtildeTodd} or the optimal observable of~\Eq{eq:ooptfano}, averaged over the sample as discussed in Sec.~\ref{sec:reco}. The predicted value of the observable is $\mathcal O=\mathcal S_{\mathcal O}\,d$, where $d$ is the coefficient of the relevant dipole operator. The response slope $\mathcal S_{\mathcal O}$ (with units of energy) is measured in simulation by extracting ${\cal O}$ from generated samples with several values of $d$ and performing a linear fit. The per-event statistical uncertainty on $\mathcal O$, denoted by $\sigma_{\cal O}$, is also measured in simulation. Specifically, $\sigma_{\cal O}$ is the standard deviation over the sample of the per-event estimator --- the quantity inside the $\langle\,\cdot\,\rangle$ in \Eq{eq:estimators} --- evaluated on the SM sample. The presence of a dipole operator can be detected at the $95\%$ CL when
$|d|>d_{95}=1.96\,\sigma_{\mathcal O}/(\mathcal S_{\mathcal O}\sqrt N)$, i.e.\ the reach on the inverse coupling is
\begin{equation}
\label{eq:reach}
\frac{1}{d_{95}}=\frac{\mathcal S_{\mathcal O}\,\sqrt N}{1.96\,\sigma_{\mathcal O}}
\qquad(\text{in TeV}).
\end{equation} 
Below, we will quote this reach for parton- and detector-level analyses at the LHC and $e^+e^-$ colliders.\footnote{Equation~\eqref{eq:reach} uses the linear response approximation. Solving the exact condition $\langle\mathcal O\rangle(d)=1.96\, \sigma_{\mathcal O}/\sqrt N$ with the full coupling dependence shifts $d_{95}$ by a relative $\mathcal O(d_{95}f_2/f_1)$ --- a fraction of a percent at the projected sensitivity. The linear truncation is also the EFT-consistent order: dipole-squared contributions are formally degenerate with uncomputed dimension-eight interference.} When comparing parton- and detector-level results, we find it convenient to quantify the dilution of the observable due to detector effects as
$D=\big(\mathcal S^{\rm det}_{\mathcal O}/\sigma^{\rm det}_{\mathcal O}\big)\big/\big(\mathcal S^{\rm parton}_{\mathcal O}/\sigma^{\rm parton}_{\mathcal O}\big)$, the per-event significance ratio. The reach of the detector-level analysis is thus reduced by a factor of $D\sqrt{\varepsilon}$ compared to the corresponding parton-level result. % 

\paragraph{Background-subtracted limit.}
Equation~\eqref{eq:reach} is the statistics-limited form of a more general
background-subtracted bound.
Every observable of \Eq{FanoCPtildeTodd} is CP-odd and therefore vanishes identically in the CP-conserving SM at the level of the production amplitude, $\mathcal O_{\rm SM}=0$. However, reconstruction and
detector effects (spin-basis smearing, kinematic-fit bias, finite acceptance) or
residual backgrounds can displace the reconstructed SM expectation to a small
nonzero value even though the underlying parton-level asymmetry is exactly zero.
A realistic limit is set against the SM expectation $\mathcal O_{\rm SM}^{\rm ref}$, obtained from Monte Carlo or data control sample. The
signal is then the background-subtracted difference $\mathcal O_{\rm data}-\mathcal
O_{\rm SM}^{\rm ref}=\mathcal S_{\mathcal O}\,d$, and the coupling is resolved at
$95\%$ CL when
\begin{equation}
\label{eq:reachbkg}
\big|\mathcal O_{\rm data}-\mathcal O_{\rm SM}^{\rm ref}\big|
> 1.96\,\sqrt{\,\sigma_{\rm stat}^2+\sigma_{\rm SM}^2+\sigma_{\rm syst}^2\,},
\qquad \sigma_{\rm stat}=\frac{\sigma_{\mathcal O}}{\sqrt N},
\end{equation}
where $\sigma_{\rm SM}$ is the uncertainty on the SM reference (from the finite
Monte-Carlo or data control sample used to fix it) and $\sigma_{\rm syst}$ collects the detector systematics. Two conditions make \Eq{eq:reachbkg} collapse to the
statistics-limited \Eq{eq:reach}: (i) the reconstructed SM reference is consistent
with zero, $\mathcal O_{\rm SM}^{\rm ref}\simeq0$, so there is no bias to subtract;
and (ii) the uncertainty on $\mathcal O_{\rm SM}^{\rm ref}\simeq0$ and the systematics are subdominant, $\sigma_{\rm
SM},\sigma_{\rm syst}\lesssim\sigma_{\rm stat}$. We explicitly verified that the condition~(i) is satisfied in our analyses at both the LHC and $e^+e^-$ colliders: the reconstructed SM samples return CP-odd observables statistically consistent with zero, so the kinematic reconstruction induces no spurious CP violation. As common in phenomenological studies, we assume that the systematic effects can be controlled so that the statistical uncertainties dominate, satisfying the condition (ii). In the future, it will be important to study the extent to which the systematic uncertainties may limit the sensitivity of these searches.

\section{Analysis: $e^+e^-$ collider}
\label{sec:ee}

\subsection{Analyzing power of observables and parton-level results}

\begin{figure}[t]
\centering
\includegraphics[width=\textwidth]{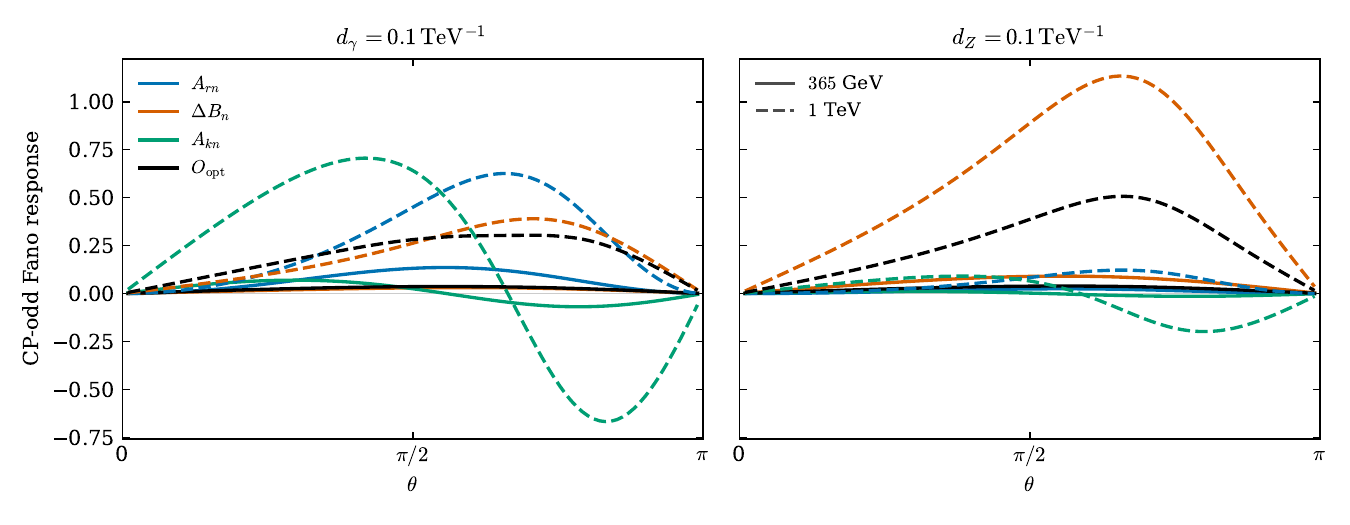}
\caption{CP-odd Fano responses ${\cal S}_{\cal O}$ versus production angle in $\eeprocess$, together with the local optimal-observable information
density \Eq{eq:oloc} (black), from the full analytic spin density matrix
evaluated for the benchmark coupling values,
$d_\gamma$ (left) and $d_Z$ (right) at $0.1\,{\rm TeV}^{-1}$ each
(normalization anchored to the
simulated response; solid: $\sqrt s=365\gev$, dashed: $1\tev$).}
\label{fig:eetheta}
\end{figure}

\begin{figure}[t]
\centering
\includegraphics[width=\textwidth]{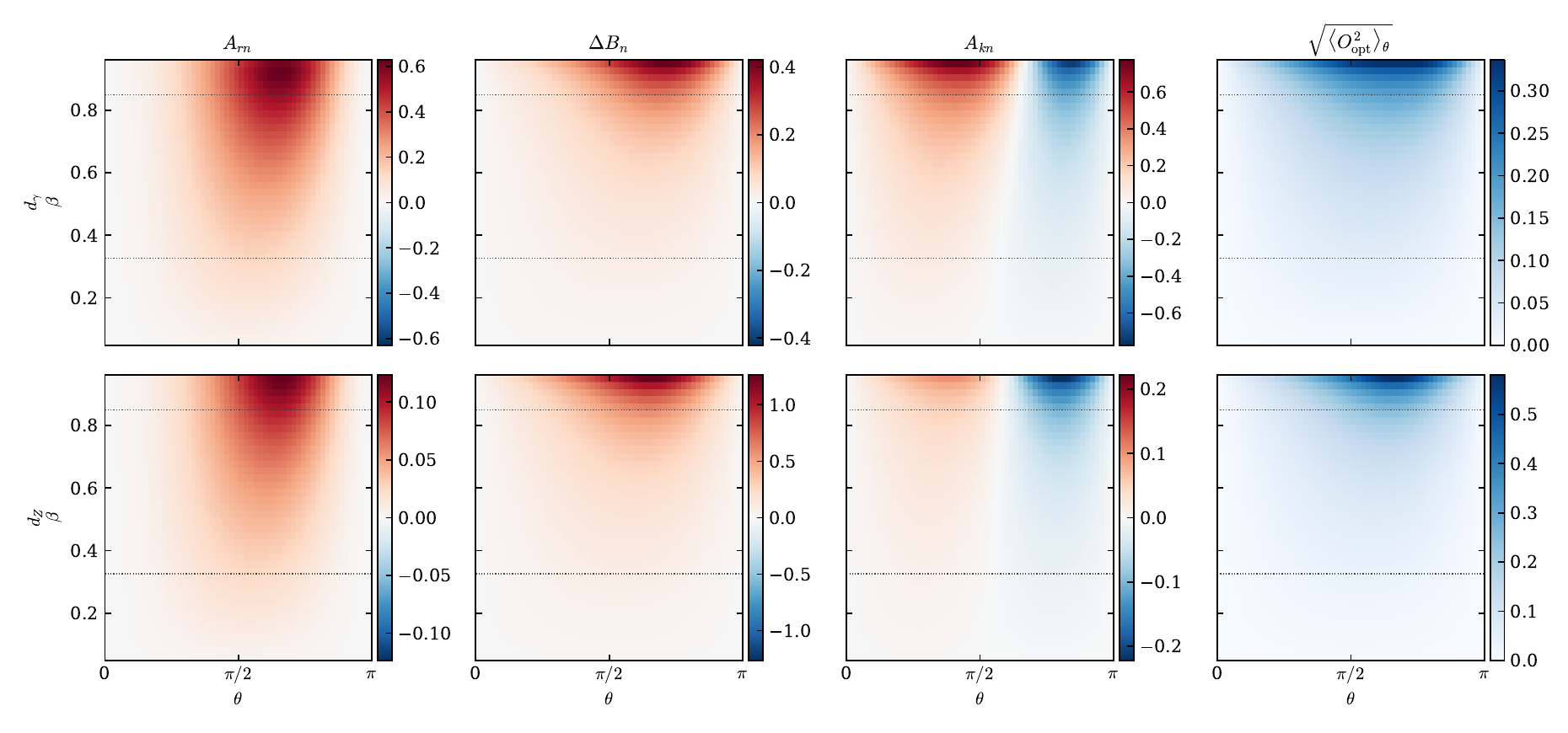}
\caption{CP-odd Fano responses ${\cal S}_{\cal O}$ and the local optimal-observable information
density \Eq{eq:oloc} over the full $(\theta,\beta)$ plane in
$\eeprocess$ ($\beta\leftrightarrow\sqrt s$; dotted lines mark the
$365\gev$ and $1\tev$ benchmarks), from the analytic spin density matrix.
Top row:
$d_\gamma=0.1\,{\rm TeV}^{-1}$, $d_Z=0$; bottom row:
$d_\gamma=0$, $d_Z=0.1\,{\rm TeV}^{-1}$.}
\label{fig:ee2d}
\end{figure}

In $\eeprocess$ the dipole signal arises from the interference of the CP-odd vertex
with the SM $\gamma^*/Z^*$ production amplitude, and the photon and $Z$ dipoles
populate the observables of \Eq{FanoCPtildeTodd} differently. The photon couples purely
vectorially (through the electric charges $Q_e,Q_t$), whereas the $Z$ couples
through both vector and axial currents, with a large electron axial charge
$g_A^e=-\tfrac12$. A nonzero single-spin normal polarization $B_n^\pm$ is T-odd and
requires an axial--vector interference, which the $Z$ supplies but the
photon does not. The two-spin correlation $A_{rn}$, by contrast, is fed efficiently
by the vector coupling. Hence $d_\gamma$ projects mainly onto $A_{rn}$ and $d_Z$
mainly onto $\Delta B_n$, while $A_{kn}$ is kinematically suppressed at threshold.
Figure~\ref{fig:eetheta} shows the analytic $\theta$-dependence of each of these observables, separately for $d_\gamma$ and $d_Z$, at $\sqrt{s}=365$~GeV (FCC-ee threshold run) and $1\tev$.
The black curves show the corresponding local quantity for the optimal
observable $O_{\rm opt}(\Omega)$ in Eq.~\leqn{eq:ooptfano}. The mean of $O_{\rm opt}(\Omega)$ at fixed $\theta$ vanishes identically, 
so the informative local scalar is its root-mean-square,
\begin{equation}
\label{eq:oloc}
\sqrt{\big\langle O_{\rm opt}^2\big\rangle_\theta}
=\left[\frac{\int d\Omega_+\,d\Omega_-\;
       O_{\rm opt}^2(\Omega)\, f_0(\Omega)}
            {\int d\Omega_+\,d\Omega_-\; f_0(\Omega)}\right]^{1/2}_{\theta\
	    {\rm fixed}}\,,
\end{equation}
with $f_0$ the SM density of \Eq{eq:oopt}. The quantity in Eq.~\leqn{eq:oloc} is the
square root of the conditional Fisher information at fixed $\theta$: by \Eq{eq:cs} it bounds the local
significance of every observable at every $\theta$, and its
rate-weighted average in quadrature is the Cram\'er--Rao
ceiling ${\rm Var}_0(O_{\rm opt})^{1/2}$. 
All considered observables grow
markedly with energy, $A_{kn}$ switches on away from threshold
with its characteristic sign change across $\theta=\pi/2$, and the
forward--backward asymmetric shapes reflect the $Z$ axial coupling. The full
$(\theta,\beta)$ dependence of the responses and of the information density
\Eq{eq:oloc} is mapped in Figure~\ref{fig:ee2d}.

\begin{table}[t!]
\centering
\begin{tabular}{lcccc}
\toprule
coupling & $A_{rn}$ & $A_{kn}$ & $\Delta B_n$ & $O_{\rm opt}$ \\
\midrule
$d_\gamma$ & $29$ & $6.1$ & $13$ & $\mathbf{42}$ \\
$d_Z$         & $5.0$ & $0.6$ & $38$ & $\mathbf{42}$ \\ %
\bottomrule
\end{tabular}
\caption{Projected parton-level $95\%$ CL reach $1/d_{95}$ (in $\tev$) on the
electric dipoles at FCC-ee/$365\gev$, with the integrated luminosity of $3\ab$. Each column is a
different CP-odd observable; the strongest reach (largest $1/d_{95}$) for each coupling
is shown in bold. 
}
\label{tab:parton}
\end{table}

We simulate $\eeprocess$ at $\sqrt s=365\gev$ (the FCC-ee threshold run,
$\beta\simeq0.33$) with $\sig_{\rm dilep}\simeq0.023$~pb and $\mathcal L=3\ab$; the
estimator r.m.s.\ read from the simulation are $\sigma_{A_{rn}}=4.2$ and
$\sigma_{\Delta B_n}=2.4$. \Tab{tab:parton} lists the resulting parton-level reach
$1/d_{95}$ from \Eq{eq:reach} for each observable--coupling pair.\footnote{For $A_{kn}$, statistically consistent with zero in the simulated
sample at threshold, the slope is instead computed from $C_{kn}-C_{nk}$ of
the inclusive density matrix \eqref{eq:rhoincl}, with the production weight
$d\sigma\propto A(\theta)$ taken from the same analytic density matrix
(Appendix~\ref{app:rho}); its estimator r.m.s.\ is
the same as for $A_{rn}$. We verified that the analytic
and simulated responses agree within statistics for $A_{rn}$ and
$\Delta B_n$, where both are available.} While $A_{kn}$ is kinematically suppressed at threshold, the other two observables provide an impressive sensitivity to the CP-odd couplings, probing scales of order 10 TeV. We also compute the reach of the optimal observable $O_{\rm opt}$ from the same sample. 
As expected, the optimal observable provides the best reach, about 40 TeV for both $d_\gamma$ and $d_Z$. 

\begin{figure}[t!]
\centering
\includegraphics[width=\textwidth]{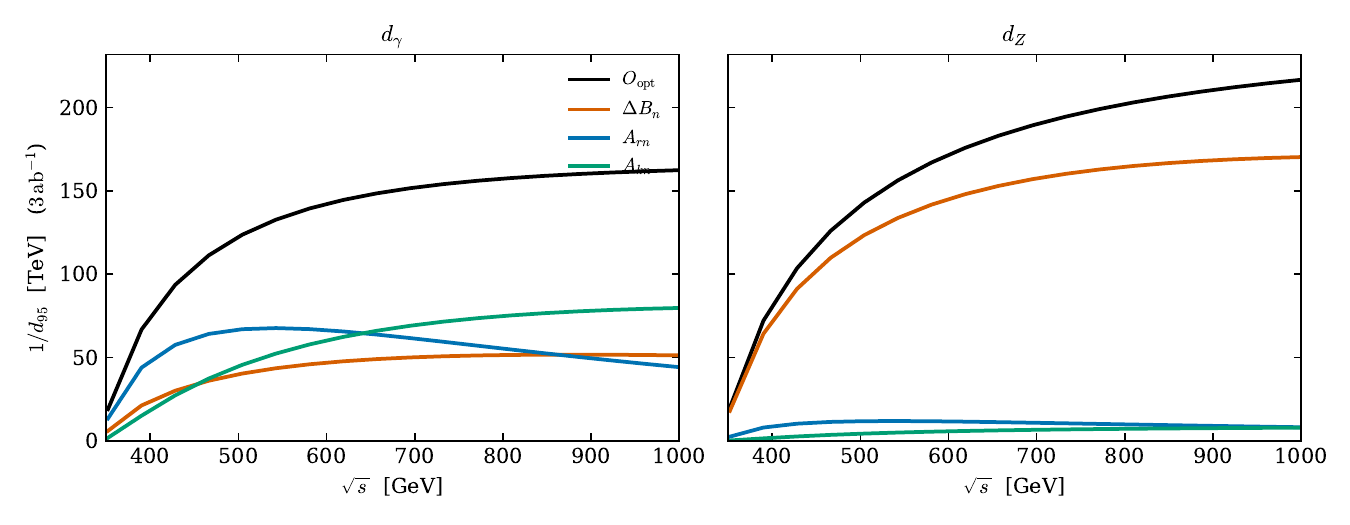}
\caption{Parton-level $95\%$ CL reach $1/d_{95}$ on $d_\gamma$ (left) and
$d_Z$ (right) versus collider energy at fixed integrated luminosity
$\mathcal L=3\ab$, obtained from the analytic spin density matrix with the
dilepton cross section normalized to the simulation at $365\gev$.
}
\label{fig:eereach}
\end{figure}

While the reach is impressive, the collider energy close to the top pair-production threshold is not optimal for this search. The contribution of the dipole operator to spin density matrix is suppressed by the momentum transfer, and hence all our observables grow with collision energy. 
Increasing the collision energy from $365\gev$ to $1\tev$, the normal correlation $A_{rn}$ nearly triples and the normal polarization $\Delta B_n$ grows eight-fold, while the $k$--$n$ correlation $A_{kn}$, which is kinematically suppressed at threshold, turns on strongly, reaching $|\langle A_{kn}\rangle|=0.49$ at $1\tev$. Folding the energy growth of the observables against the falling cross section
at fixed luminosity, Figure~\ref{fig:eereach} shows the resulting parton-level
$95\%$ CL reach as a function of the collider energy. The reach of the
optimal observable grows monotonically over the energies studied, reaching
about four ($d_\gamma$) to five ($d_Z$) times the threshold-run value at
$\sqrt s=1\tev$. 
A moderate- to high-energy $e^+e^-$ run 
(e.g. the ILC~\cite{ILC:2013jhg,ILCInternationalDevelopmentTeam:2022izu} at $\sqrt s\gtrsim0.5\tev$) thus provides the ideal setting to search for CP-violating dipole operators using quantum tomography. In contrast, the FCC-ee threshold region is optimal for measurement of quantum information observables such as entanglement and discord, since the top pair is maximally entangled at threshold. 

\subsection{Detector-level results}
\label{sec:detector}
\label{sec:results}

We now pass the simulated FCC-ee samples through showering simulation, the \texttt{IDEA} detector
simulation, and the kinematic fit of \Sec{sec:reco}, and quantify the
reconstruction efficiency and the dilution of each observable.

A complementary, entirely reconstruction-free CP-odd observable available
in $e^+e^-$ collisions is the lepton--beam triple product~\cite{Atwood:1991ka,Bernreuther:1993hq}
\begin{equation}
\label{eq:triple}
T=\hat p_{e^-}\cdot(\hat p_{\ell^+}\times\hat p_{\ell^-}),
\end{equation}
where $\hat p_{e^-}$ is the incoming electron beam direction, and $\hat p_{\ell^\pm}$ are the directions of the final state leptons in the lab frame. 
This observable is CP- and T-odd, and hence sensitive to the electric dipole at tree level. It does not require any information regarding $b$-jet momenta, and no fit for neutrino momenta is needed. Below, we will evaluate the sensitivity of $T$ to the electric dipole operators using the same simulated samples, and compare it to that of the tomographic observables advocated in this paper.

\paragraph{Basis choice and dilution.}
Near the $t\bar t$ production threshold 
the tops move slowly in the lab frame and their
direction is poorly measured
($\sim\!40^\circ$ resolution), so the helicity basis \Eq{eq:krn}, which requires
$\khat$, is badly degraded. The beam basis (fixed lab axes), which needs
only the top rest-frame boost and not the top direction, is far more robust:
the reconstructed spin correlation matrix retains $\|C_{\rm reco}\|/\|C_{\rm
truth}\|\simeq0.9$, where $\|C\|=(\sum_{ij}C_{ij}^2)^{1/2}$ is the Frobenius
norm, versus $\lsim0.3$ in the helicity basis. (Both numbers quoted above were
\emph{measured} in our detector simulation: the
$\sim\!40^\circ$ figure is the median angle between the fitted and true top
directions, and the $\|C\|$ ratios are computed using truth/reconstruction values on the same simulated events.)

\begin{table}[t!]
\centering
\begin{tabular}{lcccc}
\toprule
observable & top rest frame? & $\varepsilon$ & dilution $D$ & degradation $1/(D\sqrt\varepsilon)$ \\
\midrule
$\Delta B_n$ (normal pol.)        & yes & $0.6$  & $0.37$ & $3.4$ \\
$A_{rn}$ (normal corr.)           & yes & $0.6$  & $0.24$  & $5.3$ \\
$O_{\rm opt}$ (optimal weight) & yes & $0.6$ & $0.40$ & $3.2$ \\
$T$ (triple product)  & no  & $0.97$ & $1.0$  & $1.0$ \\
\bottomrule
\end{tabular}
\caption{Detector-level performance at FCC-ee/$365\gev$ (IDEA, dilepton channel).
The top-direction resolution near threshold is $\sim\!40^\circ$. The degradation
of the $95\%$ CL reach relative to parton level is $1/(D\sqrt\varepsilon)$: both
rest-frame observables pay a factor of a few, while the triple product is
unaffected.} 
\label{tab:reco}
\end{table}

Two effects degrade the reach relative to parton level:
the reconstruction efficiency $\varepsilon$ (requiring two leptons, two $b$-jets
and a valid fit for two neutrino momenta) and the dilution $D$ of the relevant observable,
$d_{95}^{\rm reco}\simeq d_{95}^{\rm parton}/(D\sqrt\varepsilon)$.
The dilution coefficients are measured on a high-statistics dipole sample ($10^5$
events).\footnote{A single sample with both electric dipoles switched on
suffices to measure the dilutions for both couplings: the dilution is a
property of the reconstruction, not of the coupling. It is the ratio of the
reconstructed to the truth-level asymmetry on the same events, and the
kinematic-fit smearing acts on the reconstructed axes and leptons in the
same way whichever coupling generated the correlation; since $d_\gamma$ and
$d_Z$ populate the same density-matrix entries ($C_{rn}-C_{nr}$ for
$A_{rn}$, $B^\pm_n$ for $\Delta B_n$), each observable carries a single
dilution at linear order, and the couplings differ only in their signal
slopes.} We find $\varepsilon\simeq0.6$ and an observable-dependent
dilution, summarized in \Tab{tab:reco}. All rest-frame observables are
substantially diluted by the poorly measured top direction, with
$D_{A_{rn}}=0.24(3)$, $D_{\Delta B_n}=0.37(2)$, and $D_{O_{\rm opt}}=0.40(5)$. For the optimal observable the dilution is measured per coupling, $D_{O_{\rm opt}}=0.36(6)$ for $d_\gamma$ and $0.44(7)$ for $d_Z$ (the value quoted above is the combined-coupling one); these per-coupling values, with the baseline efficiency $\varepsilon=0.62$, are what enter the $O_{\rm opt}$ row of \Tab{tab:reach}. % 
The normal polarization is in fact somewhat more robust than the normal correlation. The correlation
$A_{kn}$, by contrast, acquires a reconstruction-induced offset rather
than a pure attenuation (its ratio and regression slope disagree), and is not
used in this analysis. (It is in any case kinematically suppressed at threshold, as discussed above.)
The reconstruction-robust triple
product \Eq{eq:triple}, which needs neither the top frame nor $b$-jets, has
$\varepsilon\simeq0.97$ and $D\simeq1$.

\begin{table}[t!]
\centering
\small
\begin{tabular}{lccccccc}
\toprule
selection & $\varepsilon$ & med.\ $\delta\theta_{\rm top}$ &
$D_{A_{rn}}$ & $D_{\Delta B_n}$ & $D_{O_{\rm opt}}$ &
$D_{A_{rn}}\sqrt\varepsilon$ & $D_{\Delta B_n}\sqrt\varepsilon$ \\
\midrule
baseline                & $0.62$ & $44^\circ$ & $0.24$ & $0.37$ & $0.40$ & $0.19$ & $0.29$ \\
$\chi^2<100$            & $0.28$ & $38^\circ$ & $0.27$ & $0.37$ & $0.49$ & $0.14$ & $0.20$ \\
$\chi^2<10$             & $0.06$ & $23^\circ$ & $0.41$ & $0.55$ & $0.74$ & $0.10$ & $0.13$ \\
$\chi^2<3$              & $0.03$ & $11^\circ$ & $0.75$ & $0.71$ & $0.72$ & $0.13$ & $0.12$ \\
$p_\ell>20\gev$         & $0.57$ & $43^\circ$ & $0.25$ & $0.37$ & $0.39$ & $0.19$ & $0.28$ \\
\bottomrule
\end{tabular}
\caption{Selection-cut scan at FCC-ee/$365\gev$. For each cut, we list the reconstruction efficiency $\varepsilon$, the 
median top-direction error $\delta\theta_{\rm top}$, and the dilution $D_{\cal O}$ for each of the observables of interest. The last two columns show the degradation of the reach for each observable, relative to the parton-level estimate.} 
\label{tab:cuts}
\end{table}

\paragraph{Selection cuts.}
Could a tighter selection reduce the dilution? \Tab{tab:cuts} scans the two
natural handles. A cut on the kinematic-fit $\chi^2$ genuinely sharpens the
reconstructed objects: the median top-direction error improves from
$44^\circ$ to $11^\circ$, and the dilutions rise accordingly, up to
$D\simeq0.7$--$0.75$ at $\chi^2<3$ for all three rest-frame observables,
$A_{rn}$, $\Delta B_n$ and $O_{\rm opt}$ alike. This demonstrates that the fit
$\chi^2$ faithfully tracks per-event reconstruction quality. The efficiency, however,
falls faster than the dilution recovers: the reach degradation
$1/(D\sqrt\varepsilon)$ \emph{worsens} at every lower $\chi^2$ working point, because
the poorly fit events, while individually degraded, still carry net CP-odd
information. A lepton-momentum threshold dependence is simply flat: up to $p_\ell>20\gev$
neither $\varepsilon$ nor $D$ moves appreciably, so a realistic trigger or
identification floor costs little (below $7\%$ in the figure of merit for every observable). We therefore keep the uncut baseline
selection.

\paragraph{SM reference.}
To check the condition~(i) for the background-subtracted limit of
\Eq{eq:reachbkg} we have passed a pure-SM sample through the same reconstruction pipeline. The
reconstructed CP-odd observables were found to be consistent with zero ($A_{rn}^{\rm SM,reco}=0.07\pm0.05$, $\Delta B_n^{\rm SM,reco}=0.01\pm0.03$, and $\langle O_{\rm opt}\rangle^{\rm
SM,reco}=0.008\pm0.007$), so
the kinematic fit induces no spurious CP-odd bias and the quoted reach is a genuine
background-subtracted $95\%$ CL. The dilution coefficients listed in \Tab{tab:reco} are evaluated using the
background-subtracted responses.

\begin{table}[t!]
\centering
\begin{tabular}{lcc}
\toprule
observable & $d_\gamma$ (photon) & $d_Z$ ($Z$) \\
\midrule
$A_{rn}$ (tomography)  & $5.4$ & $1.2$ \\
$\Delta B_n$ (tomography) & $3.8$ & $11$ \\
$O_{\rm opt}$ (optimal weight) & $\mathbf{12}$ & $\mathbf{15}$ \\
$T$ (robust) & \multicolumn{2}{c}{$10$ \ (combined $d_\gamma\!=\!d_Z$)} \\
\bottomrule
\end{tabular}
\caption{Projected detector-level $95\%$ CL reach $1/d_{95}$ (in $\tev$, for ${\cal L}=3\ab$) for
the electric dipoles at FCC-ee/$365\gev$. The strongest reach for each coupling is in
bold.}
\label{tab:reach}
\end{table}

\paragraph{Results.}
\Tab{tab:reach} combines the parton-level reach of \Tab{tab:parton} with the
reconstruction degradation of \Tab{tab:reco} to give the projected detector-level
reach at FCC-ee/$365\gev$. Both tomographic observables survive reconstruction,
and their parton-level complementarity persists: $A_{rn}$ is more sensitive to $d_\gamma$ and
$\Delta B_n$ to $d_Z$. The optimal weight $O_{\rm opt}$, measured on the same reconstructed events with
its per-coupling dilutions, provides the best sensitivity. In particular, we find that the sensitivity of $O_{\rm opt}$ slightly exceeds that of the conventional triple-product $T$. Thus, our analysis underlines the value of the quantum tomography approach in the search for CP-violating effects in $t\bar{t}$. It also highlights the complementarity between the tomographic and conventional observables: while the former have significantly higher parton-level statistical sensitivity, the latter are less sensitive to detector effects. In summary, a tomography-based search at the FCC-ee will be able to probe new physics in the electric dipole couplings at the scale of around 10 TeV, well beyond the reach of current LHC measurements of these operators (see e.g.\ Ref.~\cite{CMS:2025dpp}).

A parton-level analysis indicates that the reach of the proposed search is maximized at collider energies $\sqrt{s}\simeq500$--$600\gev$; see Fig.~\ref{fig:eereach}. At the detector level, there is an additional factor which favors higher energies for tomography-based analysis. As mentioned above, the accuracy of reconstructed top direction is poor near the top production threshold, where tops are non-relativistic. At higher energy, the reconstruction improves rapidly, and the dilution of the observables is reduced. We leave a detailed study of this effect for future work.

\section{Analysis: the LHC}
\label{sec:lhc}

\begin{figure}[t!]
\centering
\includegraphics[width=0.62\textwidth]{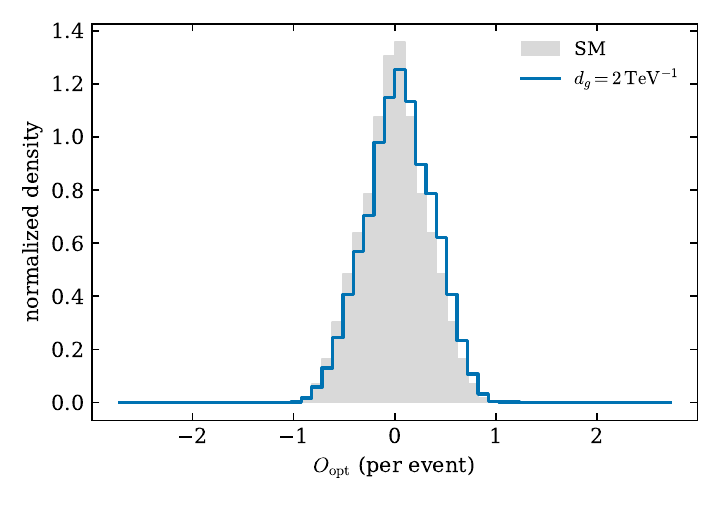}
\caption{Distribution of the hadron-level per-event optimal observable at the LHC, in the SM and with the chromo-electric dipole switched on. As in \Fig{fig:ooptdist}, the shift of the distribution to the right is the signal of CP-violation; the illustrative coupling
$d_g=2\,{\rm TeV}^{-1}$ makes the skew visible by eye.}
\label{fig:lhcooptdist}
\end{figure}

In this section we consider the process $pp\to t\bar{t}$ at the LHC. We focus on the chromo-electric dipole operator in Eq.~\leqn{eq:Lqcd}, whose contribution is enhanced due to the strong coupling of the gluon to the initial state. As before, the tomographic observables\footnote{We continue to use the helicity basis shown in Fig.~\ref{fig:axes} at the LHC. In $pp$ collisions, the transverse axes $\hat{r}$ and $\hat{n}$ are defined with an extra factor $\mathrm{sgn}(\cos\theta)$, i.e.\ $\rhat=\mathrm{sgn}(\cos\theta)\,(\hat p-\cos\theta\,\khat)/\sin\theta$, which reflects the symmetry of the initial state under $z\to-z$. This is the standard convention of Ref.~\cite{Bernreuther:2015yna}, also used in the ATLAS/CMS spin-correlation measurements.} potentially sensitive to CP-violation are $A_{rn}$, $A_{kn}$, and $\Delta B_n$. Since QCD interactions are parity-invariant, the polarizations $B^\pm_n$ vanish up to subdominant electroweak contributions. We will therefore not consider $\Delta B_n$ in the analysis of this section. The optimal tomographic observable is again given by Eq.~\leqn{eq:ooptfano} (with $B_n^\pm=0$). Note that the coefficients $f_{0,1}$ in the definition of $O_{\rm opt}$ (or equivalently the corresponding Fano coefficients) are computed separately for the $q\bar{q}$ and $gg$ production channels, defining a separate optimal observable for each partonic channel. To construct the hadron-level version of the optimal observable, we average the parton-level $f_{0,1}$ coefficients with weights proportional to the differential luminosity of the corresponding parton channel evaluated at the event's kinematics, $(m_{t\bar t},\cos\theta,\hat q_\pm)$. At the LHC, the overall top pair production is dominated by the $gg$ channel, which contributes $80-90$\% of the rate for $m_{t{bar t}}=340$~GeV$\ldots 1$~TeV. Hence the optimal observable follows closely the parton-level definition in the $gg$ channel in most of the parameter space. The distribution of the hadron-level $O_{\rm opt}$ at the LHC, in the SM and in the presence of a chromo-electric dipole, is shown in Figure~\ref{fig:lhcooptdist}.        

\subsection{Analyzing power of observables and parton-level results}

\begin{figure}[t!]
\centering
\includegraphics[width=\textwidth]{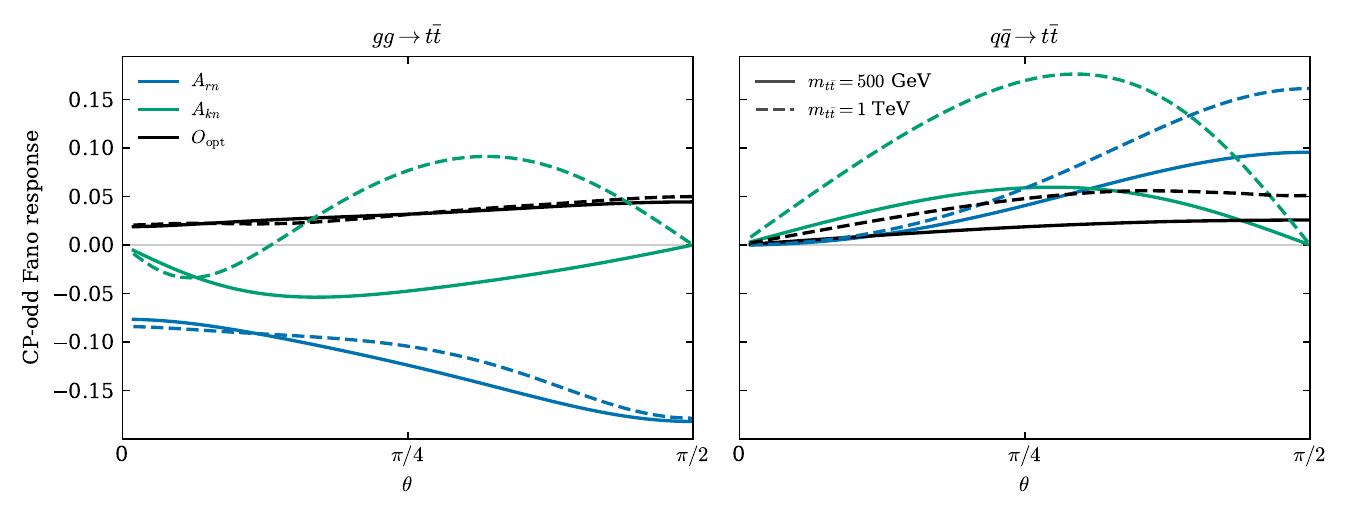}
\caption{CP-odd Fano responses ${\cal S}_{\cal O}$ versus parton-frame production angle at the LHC, together
with the local optimal-observable information density \Eq{eq:oloc} (black),
in the $gg$ (left) and $q\bar q$ (right) production
channels, %from the full analytic spin density matrix at
at the benchmark value $d_g=0.1\,{\rm TeV}^{-1}$ (solid: $m_{t\bar t}=500\gev$, dashed: $1\tev$).}
\label{fig:lhctheta}
\end{figure}

\begin{figure}[t]
\centering
\includegraphics[width=\textwidth]{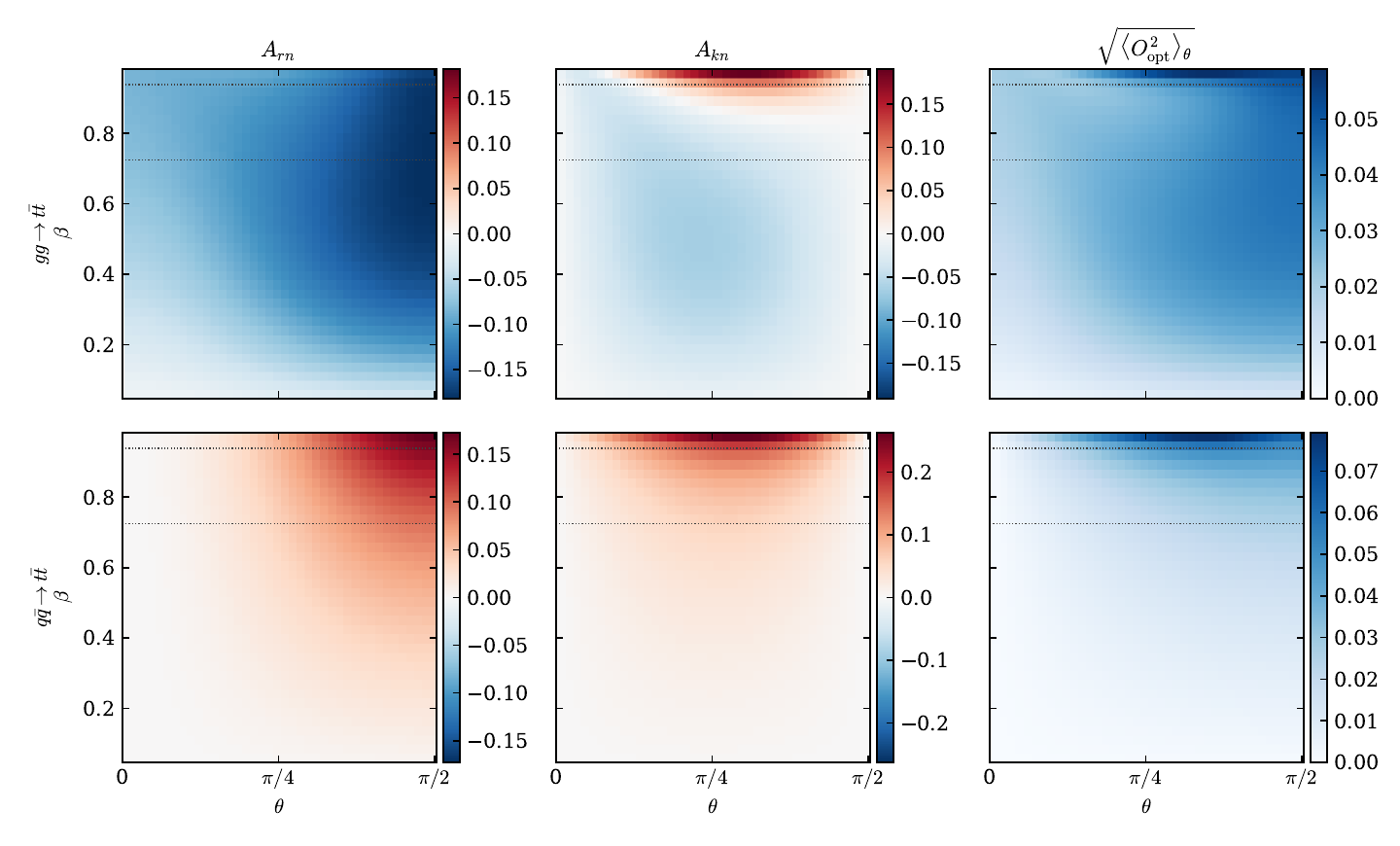}
\caption{CP-odd Fano responses ${\cal S}_{\cal O}$ and the local optimal-observable information
density \Eq{eq:oloc} (rightmost column) over the $(\theta,\beta)$ plane at
the LHC, for the $gg$ (top) and $q\bar q$ (bottom) channels at
$d_g=0.1\,{\rm TeV}^{-1}$ (dotted
lines mark $m_{t\bar t}=500\gev$ and $1\tev$).}
\label{fig:lhc2d}
\end{figure}

Figures~\ref{fig:lhctheta} and~\ref{fig:lhc2d} map the parton-level
analyzing power, in exact analogy with Figures~\ref{fig:eetheta}
and~\ref{fig:ee2d} for the $e^+e^-$ collider. Note that the production angle is now restricted to $\theta\in[0,\pi/2]$ due to the $z\to -z$ symmetry of the $pp$ initial state. The curves for $A_{rn}$ and $A_{kn}$ were computed both using the analytic form of the Fano coefficients and from the Monte Carlo simulation; the agreement between the two methods serves as a validation of the MC samples. Note that in most cases, the responses ${\cal S}_{\cal O}$ of the two parton-level channels, $gg$ and $q\bar{q}$, have opposite signs, leading to partial cancellation in the analyzing power. However, since the $gg$ channel strongly dominates the net signal ($87\%$ of the dileptonic sample at $13\tev$), the combined reach does not suffer significantly.  

\begin{figure}[t!]
\centering
\includegraphics[width=0.62\textwidth]{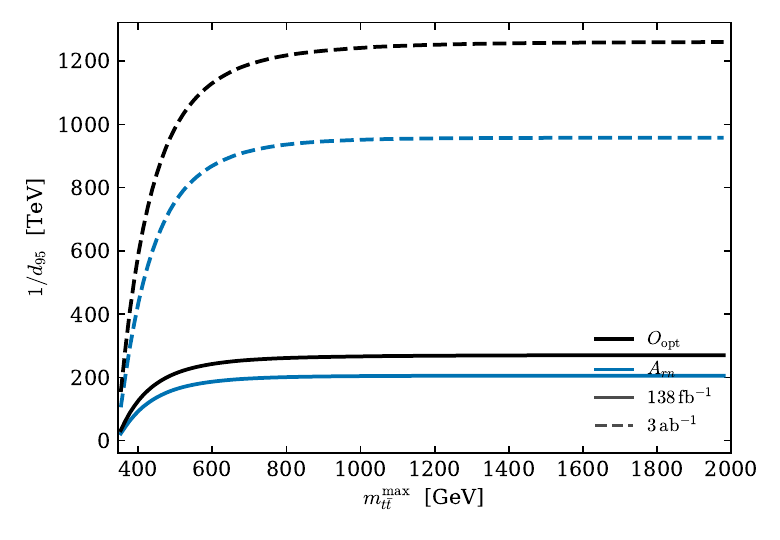}
\caption{Cumulative
parton-level $95\%$ CL reach $1/d_{95}$ versus an upper $m_{t\bar t}$ cut,
with the $gg$ and $q\bar q$ channels mixed according to the simulated SM
sample, at $138\,{\rm fb}^{-1}$ (solid) and $3\ab$ (dashed), using the
simulated dileptonic cross section $\sigma=20.4$~pb at $\sqrt s=13\tev$.
As in \Fig{fig:eereach} these are parton-level idealizations (perfect
reconstruction, no backgrounds); the corresponding detector-level reaches
are given in \Tab{tab:lhcreach}.}
\label{fig:lhcreach}
\end{figure}

The parton-level estimate of the reach of the LHC for the chromo-electric dipole operator is shown in Figure~\ref{fig:lhcreach}. For illustration, we show the reach as function of an upper cut on the $t\bar{t}$ invariant mass $m_{t\bar{t}}$. The parton-level analyzing power of the observables grows rapidly with energy, but a fast fall-off of the parton distribution functions leads the yield to saturate above $m_{t\bar t}\simeq 600$~GeV. The cumulative reach is quite impressive, extending up to about $1/d_g\approx 270$~TeV for the LHC Run-2 dataset (${\cal L}_{\rm int}=138$~fb$^{-1}$) and to $1/d_g\approx 1260$~TeV for the HL-LHC (${\cal L}_{\rm int}=3$~ab$^{-1}$). We do not plot the limit provided by the observable $A_{kn}$ alone, since it is significantly weaker than $A_{rn}$; however, it does contribute to the improved reach of the optimal observable compared to $A_{rn}$, as is clear from the figure.

\subsection{Detector-level results}

To obtain a more realistic estimate of the LHC reach, we performed a parametrized detector-level analysis. The Monte Carlo sample for this analysis was obtained by showering the parton-level dileptonic top samples with \texttt{Pythia8}~\cite{Sjostrand:2014zea} and passing them
through a fast simulation of the CMS detector (\texttt{Delphes})~\cite{deFavereau:2013fsa}. We then applied a st of realistic CMS-like selection criteria: two opposite-sign leptons
with $p_T>25/20\gev$ and $|\eta|<2.4$, $m_{\ell\ell}>20\gev$, a $Z$ veto
$|m_{\ell\ell}-m_Z|>15\gev$ plus $\slashed E_T>40\gev$ in the same-flavor
channels, and at least two jets ($p_T>30\gev$, $|\eta|<2.4$) of which two are $b$-tagged (we also consider a single-$b$-tag category below). We then applied the top reconstruction algorithm of Section~\ref{sec:decaymethod}. As explained in that section, due to lack of knowledge of production frame and longitudinal momentum, the fit has less inputs at the LHC than in the $e^+e^-$ case, and the jet scale factors cannot be fixed by the fit. This leads to poorer top momentum reconstruction overall. At the same time, the tops at the LHC are more likely to be produced well above threshold, with relativistic velocities in the lab frame, than at the FCC-ee. This makes the top direction in the lab frame easier to reconstruct precisely.

In the analysis of $e^+e^-$ case in Section~\ref{sec:ee}, we found it useful to compare the tomographic observables studied here with the more conventional CP-odd triple-product observable, which does not require jet momenta measurement or top reconstruction. Unfortunately, this observable is not available at the LHC, because it is odd under the beam flip $z\to-z$ while the LHC $pp$ initial state is symmetric. Instead, the
CP-odd, Lorentz-invariant quadruple product observable
\begin{equation}
\label{eq:O1}
O_1=\epsilon_{\mu\nu\rho\sigma}\,p_t^{\mu}\,p_{\bar t}^{\nu}\,p_{\ell^+}^{\rho}\,p_{\ell^-}^{\sigma},
\end{equation}
was proposed in Ref.~\cite{Hayreter:2015ryk} and used in the CMS CEDM
search~\cite{CMS:2022quh}. A non-zero sample-mean value of $O_1$ indicates the presence of non-SM CP-violating physics.\footnote{In practice, CMS used the sign asymmetry of $O_1$ rather than the mean, since the mean is
dominated by heavy kinematic-fit tails; we use the same procedure in our analysis of $O_1$.} Unlike the triple-product, $O_1$ does require   
top momentum reconstruction, and thus is affected by some of the same detector effects as the tomographic observables. In our detector analysis, we will 
evaluate the sensitivity of $O_1$ and the tomographic observables using the same MC samples and top reconstruction algorithm, and compare them directly. As an additional check of our procedure, we verified that our detector-level analysis approximately reproduces the $O_1$ exclusion reported by CMS~\cite{CMS:2022quh} after appropriate luminosity rescaling (see below for details).

Just as in the $e^+e^-$ analysis above, we verified that the mean value of all CP-odd observables of interest is consistent with zero when evaluated on SM-only samples, generated and analyzed with the same procedures as our main analysis. (For example, we measure $A_{rn}^{\rm SM,reco}=-0.03\pm0.04$.) We conclude that acceptance and selection cuts, as well as the neutrino-weighting fit used for top momentum reconstruction, do not induce any spurious CP violation.

\begin{table}[t!]
\centering
\begin{tabular}{lccccc}
\toprule
selection & $\varepsilon$ & $D_{A_{rn}}$ & $D_{O_{\rm opt}}$ &
$D_{A_{rn}}\sqrt\varepsilon$ & $D_{O_{\rm opt}}\sqrt\varepsilon$ \\
\midrule
objects only                 & $0.109$ & $0.53(6)$ & $0.50$ & $0.175$ & $0.165$ \\
baseline                     & $0.068$ & $0.57(6)$ & $0.58$ & $0.147$ & $0.150$ \\
baseline, $e\mu$ only        & $0.042$ & $0.63(10)$ & --- & $0.128$ & --- \\
baseline $+\chi^2\!<\!20$    & $0.044$ & $0.61(7)$ & $0.56$ & $0.127$ & $0.117$ \\
baseline $+$ 1-$b$-tag cat.  & $0.157$ & $0.42(4)$ & $0.50$ & $0.168$ & $0.199$ \\
\midrule
$m_{t\bar t}<450\gev$                          & --- & $0.33(14)$ & --- & --- & --- \\
$m_{t\bar t}\in[600,800]\gev$                  & --- & $0.69(15)$ & --- & --- & --- \\
$m_{t\bar t}\in[800,1200]\gev$                 & --- & $0.71$     & --- & --- & --- \\
\bottomrule
\end{tabular}
\caption{LHC selection scan (Delphes CMS, dilepton). ``Objects only'' requires two opposite-sign leptons and two $b$-tagged jets; ``baseline'' adds the trigger thresholds ($p_T>25/20\gev$, $|\eta|<2.4$), $m_{\ell\ell}>20\gev$ and the same-flavor $Z$-veto$+\slashed E_T$ cuts. The last row adds the single-$b$-tag category ($b$-tag plus hardest untagged jet). Top block: reconstruction efficiency, dilutions, and the reach figures of merit $D\sqrt\varepsilon$ for $A_{rn}$ and $O_{\rm opt}$ (larger f.o.m. corresponds to higher reach). Bottom block: dilution binned in $m_{t\bar t}$ under the baseline selection criteria.}
\label{tab:lhccuts}
\end{table}

\paragraph{Dilution and selection cuts.} The detector-level reach is affected by the reconstruction efficiency $\varepsilon$ and dilution $D$, with the overall degradation of the reach by a factor of $D\sqrt{\varepsilon}$ relative to the parton-level estimate. To optimize the detector-level reach, we studied a variety of selection-cut choices beyond the baseline selections listed above. The results are summarized in 
\Tab{tab:lhccuts}. The efficiency of the CMS-like baseline is $\varepsilon\simeq0.068$. As expected, this is a factor $\sim\!10$ below the FCC-ee efficiency, owing to higher-multiplicity final states, trigger thresholds, and the $b$-tagging requirement. The $e\mu$ channel, free of the $Z$-veto and $\slashed E_T$ cuts, carries most of the sensitivity. However, dilution coefficients are somewhat better than at the FCC-ee due to more relativistic tops. As at the FCC-ee, a kinematic-fit quality cut or harder lepton thresholds raise $D$ slightly, but only at the expense of steep losses in efficiency, reducing the overall reach. On the other hand, adding the single-$b$-tag category (one $b$-tag plus the hardest untagged jet) more than doubles the sample size, and while the dilution is somewhat stronger than for the baseline sample, the overall reach is increased. We therefore include this category in the reach estimates below. To understand the physical origin of the dilution better, we binned the events in $m_{t\bar t}$; see the last three rows of \Tab{tab:lhccuts}. As expected, the accuracy of top direction reconstruction plays the major role here: the top-direction error in the $t\bar t$ frame falls from $34^\circ$ near threshold to $8^\circ$ at $m_{t\bar t}\sim1\tev$, with $D$ rising from $0.33$ to $\sim0.75$. In the detector-level reach calculation, the dilution factors are convoluted with the Fano responses (which also grow with $m_{t\bar t}$) and the parton-level cross sections (decreasing with $m_{t\bar{t}}$ due to falling p.d.f.'s).

\begin{table}[t!]
\centering
\begin{tabular}{lcc}
\toprule
observable & Run~2 ($138\fb$) & HL-LHC ($3\ab$) \\
\midrule
$A_{rn}$ (tomography)            & $36$ & $168$ \\
${\rm sgn}\,O_1$ (quadruple product) & $33$ & $152$ \\
$O_{\rm opt}$ (optimal weight)   & $\mathbf{41}$ & $\mathbf{191}$ \\
\bottomrule
\end{tabular}
\caption{Projected detector-level $95\%$ CL reach $1/d_{95}$ (in $\tev$) on the
chromo-EDM at the LHC, from \Eq{eq:reach} with the selection of
\Tab{tab:lhccuts} (best reach in bold). The projected reach for ${\rm sgn}\,O_1$ approximately agrees with the result reported by  CMS~\cite{CMS:2022quh} after luminosity rescaling.}
\label{tab:lhcreach}
\end{table}

\paragraph{Results.} Projected detector-level reach of the tomographic observables at the LHC Run 2 and HL-LHC is shown in \Tab{tab:lhcreach}. Based on the above discussion, we choose the combination of selection criteria which minimizes the reach degradation relative to parton-level reach, namely the baseline selection cuts supplemented with the single-$b$-tag sample. For 
comparison, we also show the estimated reach of the conventional observable $O_1$, currently used by the CMS search for chromo-electric top dipole. The ${\rm sgn}\,O_1$ and $O_{\rm opt}$ rows are obtained from the $A_{rn}$ reach by rescaling with the per-event significance ratios measured on the same selected reconstructed sample, i.e.\ by their significance dilutions. %
A direct comparison with the published CMS result validates our $O_1$
projection. Rescaled to the $35.9\fb$ dataset of Ref.~\cite{CMS:2022quh},
our statistics-only ${\rm sgn}\,O_1$ reach corresponds to
$1/d_g\simeq17\tev$ ($\mathrm{Im}\,c_{tG}\lesssim0.24$), in good agreement
with the observed CMS limit, $\mathrm{Im}(d_{tG})\in[-0.3,0.3]$, i.e.\
$1/d_g\simeq14\tev$. The remaining $\sim20\%$ difference is consistent with
the backgrounds and experimental systematics that our projection neglects.

Thanks to high statistics and large strong-interaction cross sections, the LHC has an impressive reach in this channel, probing CP-violating new physics at scales up to about $40$~TeV (Run 2) and $200$~TeV (HL-LHC). This is significantly above the scales probed by the electroweak dipole operators at either the LHC or the FCC-ee, although of course the comparison between different operators is intrinsically model-dependent, and both searches are well worth pursuing. Our analysis indicates that the tomographic observables, $A_{rn}$ and $O_{\rm opt}$, are somewhat more sensitive than the conventional $O_1$. We hope that this motivates the LHC experiments to include tomographic observables in the future searches for CP-violating physics in top pair-production.

\section{Conclusions}
\label{sec:conclusions}

In this paper we proposed novel observables that can be used to search for CP-violating top quark couplings at both $e^+e^-$ and hadron
colliders. The proposed observables are based on spin-density-matrix tomography. Based on symmetry arguments alone, we identified the components of the spin-density matrix that carry information about CP violation in the top sector. Under very general conditions, the three relevant observables are the normal-polarization asymmetry $\Delta B_n$, and two antisymmetric normal-plane correlations $A_{rn}$, $A_{kn}$. Each of these observables can be used independently as a probe of CP-violating top couplings. We also showed how they can be combined with kinematic information (top decay product directions in the lab frame) to form the optimal observable $O_{\rm opt}$, which provides the maximal possible statistical sensitivity to CP violation at the parton level.   

The results reported in this paper range from theory-level to simplified detector simulation-level studies. On the theory side, we computed the $\ttbar$ production spin density matrix analytically for $e^+e^-\to\ttbar$ with electroweak dipoles and for $q\bar q,gg\to\ttbar$ with the chromo-dipole operators. We present explicit formulas for the spin density matrix in the appendix, separating the Standard Model, SM-dipole interference, and 
dipole-squared orders. Further, we implemented the relevant dipole operators in the {\tt MadGraph5} suite of numerical tools, and verified this implementation against our analytic results. The numerical implementation was then used to provide parton-level and detector-level (with fast/parametrized detector simulation by {\tt Delphes}) Monte Carlo samples of top-pair events in the dilepton channel at both $e^+e^-$ and $pp$ colliders. The statistical power of the proposed observables was investigated using these samples at both the LHC (including Run 2 and HL-LHC data sets) and at a proposed future electron-positron collider (including the FCC-ee operating at $\sqrt{s}=365$~GeV, and projections for higher energy scenarios up to $\sqrt{s}=1$~TeV).

Our results indicate that the novel CP-odd observables provided by the tomographic measurement of the spin density matrix have an impressive power to search for CP violation in realistic experimental settings. Both at the LHC and the FCC-ee, the new physics reach of the tomographic observables exceeds that of the traditional observables currently used in the searches for these operators. This is true even after the dilution of the tomographic observables by detector effects (which is particularly relevant at the FCC-ee where the near-threshold kinematics makes it difficult to reconstruct the top rest frame precisely) is taken into account. We urge the experimental collaborations to incorporate the observables proposed here in future searches for CP violation in the top sector.   

In recent years, discussions of the top spin-density matrix in both theoretical and experimental literature largely focused on extracting quantum-information observables such as entanglement, discord, {\it etc.} Our proposed observables are rooted in the same tomographic measurements, but they do not have an obvious quantum-information interpretation, being instead motivated by a specific particle-physics target: CP violation. Along the way, we demonstrated that the discord asymmetry, proposed recently as a probe of CP violation, in fact vanishes nearly identically in realistic processes. This is not an accident, but a consequence of a more general no-go result we proved for a broad set of related QI observables. This highlights the fact that the tomography of the spin density matrix provides a wealth of information, and a variety of observables can be constructed out of this matrix to specifically target various physical questions. A broad exploration of the power of tomographic observables, beyond the quantum information program alone, is well motivated. We hope to continue this exploration in future work.

\paragraph{Note added.}
While this work was being completed, Refs.~\cite{Lamba:2026jfm,Lamba:2026sni}
appeared, which also study CP violation in the $\ttbar$ system through the
production spin density matrix and its tomographic reconstruction. Where the
two analyses overlap --- the closed-form Fano coefficients and the
identification of the polarization difference and the antisymmetric spin
correlations as the CP-odd structures --- the results agree. The
treatments are complementary: Refs.~\cite{Lamba:2026jfm,Lamba:2026sni}
cover a broader set of operators, additional production channels and
CP violation in the decay vertex, while the analysis presented here
constructs the statistically optimal CP-odd observable and carries the
projections to detector level. We also note that their conclusions on
quantum-information measures as CP probes are consistent with the no-go
result of \Sec{sec:qinfo}.

\acknowledgments
The authors thank Jay Hubisz for helpful discussions and feedback.
The authors are funded by the NSF grant PHY-2309456. TY is also supported by the Samsung Science and Technology Foundation under Project Number SSTF-BA2201-06.

\appendix
\section{The optimal observable from the spin density matrix}
\label{app:oopt}

\paragraph{Derivation of the optimal weight.}
The joint decay-angle distribution follows from the production spin density
matrix \Eq{eq:rho} and the spin-analyzer factorization of \Sec{sec:decaymethod}:
projecting $\rho(\cos\theta)$ onto the dilepton analyzers ($\alpha_\ell=1$)
gives the standard trilinear form~\cite{Mahlon:2010gw,Bernreuther:2015yna}
\begin{equation}
\label{eq:trilinear}
\frac{d\sigma}{d\cos\theta\,d\Omega_+ d\Omega_-}
=\frac{A(\cos\theta)}{(4\pi)^2}
\Big[1+\vec B^{+}\!\cdot\hat q_+ +\vec B^{-}\!\cdot\hat q_-
+\hat q_+\!\cdot C\,\hat q_-\Big],
\end{equation}
with all coefficients functions of $\cos\theta$ (and, at a hadron collider, of
$m_{t\bar t}$ and the initial-state mixture). Expanding each coefficient to
linear order in a dipole coupling, $X=X_0+d\,X_1+\mathcal O(d^2)$ for
$X\in\{A,\vec B^\pm,C\}$, and identifying $f_{0,1}$ of \Eq{eq:oopt} as the
corresponding orders of \Eq{eq:trilinear}, one finds
$f_1/f_0=\big[\tfrac{A_1}{A_0}\big(1+\vec B_0^+\!\cdot\hat q_+ +\dots\big)
+\vec B_1^+\!\cdot\hat q_+ +\vec B_1^-\!\cdot\hat q_-
+\hat q_+\!\cdot C_1\hat q_-\big]/\big[1+\vec B_0^+\!\cdot\hat q_+ +\dots\big]$.
For a CP-odd coupling the linear interference does not modify the
spin-summed rate, $A_1=0$ (verified explicitly in our closed forms), and the
expression reduces to \Eq{eq:ooptfano}. The construction
$d\sigma_1/d\sigma_0$ was introduced for this process in
Ref.~\cite{Atwood:1991ka} and developed into per-coupling optimal observables
for the top dipoles in Refs.~\cite{Bartl:1998nn,Durieux:2018tev}.

\paragraph{Proof of optimality.}
Write the normalized event density at linear order in a CP-odd coupling $d$ as
$p(\Omega|d)=\big[f_0(\Omega)+d\,f_1(\Omega)\big]/\sigma(d)$ with
$\sigma(d)=\sigma_0+d\,\sigma_1$, $\sigma_i=\int\!f_i\,d\Omega$, and let
$\langle\,\cdot\,\rangle_d$ denote the mean over $p(\Omega|d)$. For an
\emph{arbitrary} per-event weight $O(\Omega)$, expanding to first order in $d$
gives
\begin{equation}
\langle O\rangle_d=\langle O\rangle_0
+d\,\Big[\big\langle O\,O_{\rm opt}\big\rangle_0
-\langle O\rangle_0\big\langle O_{\rm opt}\big\rangle_0\Big]
+\mathcal O(d^2)
=\langle O\rangle_0+d\,{\rm Cov}_0\big(O,O_{\rm opt}\big),
\end{equation}
where $O_{\rm opt}=f_1/f_0$ as in \Eq{eq:oopt} and we used
$\sigma_1/\sigma_0=\langle O_{\rm opt}\rangle_0$ (for a CP-odd coupling
$\sigma_1=0$, so $\langle O_{\rm opt}\rangle_0=0$). The slope of any observable
is therefore its covariance with $O_{\rm opt}$, while its statistical noise per
event is $\sqrt{{\rm Var}_0(O)}$, so the per-event significance obeys
\begin{equation}
\label{eq:cs}
\frac{\mathcal S_O}{\sigma_O}
=\frac{{\rm Cov}_0(O,O_{\rm opt})}{\sqrt{{\rm Var}_0(O)}}
={\rm corr}_0\big(O,O_{\rm opt}\big)\,\sqrt{{\rm Var}_0(O_{\rm opt})}
\;\le\;\sqrt{{\rm Var}_0(O_{\rm opt})},
\end{equation}
by the Cauchy--Schwarz inequality, with equality iff $O$ is (affinely)
proportional to $O_{\rm opt}$. Hence $O_{\rm opt}$ is the most sensitive
per-event weight, and its significance $\sqrt{{\rm Var}_0(O_{\rm opt})}$ is the
square root of the linear-order Fisher information --- the Cram\'er--Rao bound
is saturated. Equation~\eqref{eq:cs} also quantifies the shortfall of any fixed
observable: \emph{the reach improvement available from full optimization
is exactly the inverse correlation}, $1/{\rm corr}_0(O,O_{\rm opt})$.

\paragraph{Correlations with Input Observables.} The correlations used in the above proof are defined as follows. 
All moments are taken in the SM: $\langle\,\cdot\,\rangle_0$ is the $d=0$ case
of the mean $\langle\,\cdot\,\rangle_d$ over $p(\Omega|d)$ defined in the
optimality proof above, i.e.\
$\langle X\rangle_0=\int X\,f_0\,d\Omega/\sigma_0$, with
${\rm Cov}_0(X,Y)=\langle XY\rangle_0-\langle X\rangle_0\langle Y\rangle_0$ and
\begin{equation}
{\rm corr}_0(O,O_{\rm opt})
=\frac{{\rm Cov}_0(O,O_{\rm opt})}
      {\sqrt{{\rm Var}_0(O)\,{\rm Var}_0(O_{\rm opt})}}\,.
\end{equation}
The ``combined'' row is the \emph{multiple correlation}: the largest
correlation attainable by any linear combination
$\alpha_1 w_{A_{rn}}+\alpha_2 w_{A_{kn}}+\alpha_3 w_{\Delta B_n}$ of the three
per-event estimators,
\begin{equation}
{\rm corr}_0^{\rm comb}
=\sqrt{\frac{\vec S^{\,T} V^{-1}\vec S}{{\rm Var}_0(O_{\rm opt})}}\,,
\qquad
S_i={\rm Cov}_0(w_i,O_{\rm opt}),\quad
V_{ij}={\rm Cov}_0(w_i,w_j)\,,
\end{equation}
which by \Eq{eq:cs} equals the ratio of the best-combination significance to
the optimal one. 

\begin{table}[t!]
\centering
\begin{tabular}{lcccc}
\toprule
& \multicolumn{2}{c}{$\sqrt s=365\gev$} & \multicolumn{2}{c}{$\sqrt s=1\tev$}\\
${\rm corr}_0(O,O_{\rm opt})$ & $d_\gamma$ & $d_Z$ & $d_\gamma$ & $d_Z$ \\
\midrule
$A_{rn}$      & $\mathbf{0.69}$ & $0.12$ & $0.32$ & $0.04$ \\
$A_{kn}$      & $0.15$ & $0.01$ & $0.54$ & $0.04$ \\
$\Delta B_n$  & $0.32$ & $\mathbf{0.90}$ & $0.37$ & $\mathbf{0.83}$ \\
combined      & $0.73$ & $0.90$ & $0.69$ & $0.83$ \\
\bottomrule
\end{tabular}
\caption{Correlation ${\rm corr}_0(O,O_{\rm opt})$ of each CP-odd observable
with the optimal observable
\Eq{eq:oopt}. The ``combined'' row is
the multiple correlation of the three estimators. 
}
\label{tab:oopt}
\end{table}

As an example, the resulting correlations for the production process $e^+e^-\to t\bar{t}$ at $\sqrt{s}=365$~GeV are listed in Table~\ref{tab:oopt}. These correlations are evaluated from the analytic spin
density matrix by direct numerical integration over $\Omega$,
and cross-checked by event-level measurements on the simulated samples described in \Sec{sec:sim}. The input observable providing the best reach for each coupling (shown in bold) is also the one most correlated to the optimal observable, as expected. The gain in sensitivity from using the $O_{\rm opt}$ relative to the input observables is inversely proportional to the correlation coefficients. At $365\gev$ the input observables capture $90\%$ ($d_Z$, via $\Delta B_n$) and $73\%$ ($d_\gamma$) of the optimal sensitivity, so full optimization would sharpen the parton-level reach by $\sim\!10\%$ and $\sim\!40\%$, respectively. At $1\tev$ the
$d_\gamma$ information spreads across $A_{rn}$, $A_{kn}$ and their angular
dependence, and the optimal weight gains a factor $\sim\!1.4$ even over the
best combination --- optimization pays more as the energy grows.

\section{Analytic form of the spin density matrix}
\label{app:rho}

In this appendix, we collect the closed-form expressions for the $\ttbar$ spin density matrix in three production channels: $e^+e^-$, $q\bar{q}$, and $gg$. We list the entries of the {\it un-normalized} spin-density matrix, 
\begin{equation}
\label{eq:rhoraw}
\tilde{\rho}=\frac14\Big[A\,\mathbb{1}\otimes\mathbb{1}
 +\sum_i AB_i^{+}\,\sigma_i\otimes\mathbb{1}
 +\sum_j AB_j^{-}\,\mathbb{1}\otimes\sigma_j
 +\sum_{ij} AC_{ij}\,\sigma_i\otimes\sigma_j\Big],
\end{equation}
where $A$ is the rate function, and $B$ and $C$ are the Fano coefficients of Eq.~\leqn{eq:rho}. All non-vanishing Fano coefficient functions are listed below for each of
the three production channels, at each of the three orders (SM, SM-dipole
interference, dipole squared) separately. Entries
related by the symmetry properties of \Sec{sec:cpstructure}
(e.g.\ $C_{nr}=-C_{rn}$ at CP-odd interference order, $\vec B^-=\vec B^+$ in
the SM) are given once; all entries not listed vanish identically. 

We use the following conventions: $\beta$ is the
top velocity in the $\ttbar$ center-of-mass frame, $\theta$ the production
angle in the same frame, and the $\{\khat,\rhat,\nhat\}$ basis and analyzer conventions are those of Secs.~\ref{sec:framework} and~\ref{sec:reco}. The expressions are written in the minimal variable set $(m_t,\beta,\gamma)$, with
$\gamma\equiv1/\sqrt{1-\beta^2}$ the top boost, eliminating the beam energy
through $E=\sqrt{\hat s}/2=\gamma\,m_t$; each numerator is then a polynomial
in $\cos\theta$, with a single power of $\sin\theta$ factored out where the
entry is odd. The dipole couplings
$a_\gamma,d_\gamma,a_Z,d_Z$ and $a_g,d_g$ are those of the
Lagrangians~\eqref{eq:Lew} and~\eqref{eq:Lqcd}, $e$ is the positron charge,
$g_s$ the strong coupling, $s_W,c_W$ the sine and cosine of the weak
mixing angle, $M_Z$ the $Z$ mass, and $Q_f$ the electric charge of fermion
$f$ in units of $e$. The Standard Model neutral-current couplings are normalized
as
\begin{equation}
g_{Vf}=\tfrac12\big(T^3_f-2Q_f s_W^2\big),\qquad
g_{Af}=\tfrac12\,T^3_f ,
\end{equation}
so that the SM $Zf\bar f$ vertex reads
$-i\,(e/s_Wc_W)\,\gamma^\mu\big(g_{Vf}-g_{Af}\gamma_5\big)$.

{\small
\subsection{$e^+e^-\to\ttbar$}
Throughout this subsection $s=4E^2$ and
$\chi\equiv s/(s-M_Z^2)$; each coefficient is organized by propagator
structure, $f_{\gamma\gamma}+\chi f_{\gamma Z}+\chi^2 f_{ZZ}$.

\paragraph{SM.}

\begin{flalign*}
A ={}& 3 Q_\ell^2 Q_t^2 e^4 \left[ 2 - \sin^2\theta \, \beta^2 \right] &\\
&{} + \chi \frac{6 Q_\ell Q_t e^4}{c_W^2 s_W^2} \left[ g_{V\ell} g_{Vt} \left( 2 - \beta^2 \sin^2\theta \right) + 2 g_{A\ell} g_{At} \, \beta \cos\theta \right] &\\
&{} + \chi^2 \frac{3 e^4}{c_W^4 s_W^4} \left[ 8 g_{A\ell} g_{At} g_{V\ell} g_{Vt} \, \beta \cos\theta + (g_{A\ell}^2 + g_{V\ell}^2) \left( g_{At}^2 (1 + \cos^2\theta)\beta^2 + g_{Vt}^2 (2 - \beta^2 \sin^2\theta) \right)   \right] &
\end{flalign*}

\begin{flalign*}
A \, B_k^+ ={}& \chi \frac{6 Q_\ell Q_t e^4}{c_W^2 s_W^2} \left[ - g_{At} g_{V\ell} (\cos^2\theta + 1)\beta - 2 g_{A\ell} g_{Vt} \cos\theta \right] &\\
&{} + \chi^2 \frac{6 e^4}{c_W^4 s_W^4} \left[ - 2 g_{A\ell} g_{V\ell} (g_{Vt}^2 + \beta^2 g_{At}^2) \cos\theta - g_{At} g_{Vt} (g_{A\ell}^2 + g_{V\ell}^2) \cdot \beta \left( 1 + \cos^2\theta \right) \right] &
\end{flalign*}

\begin{flalign*}
A \, B_r^+ ={}& \chi \left( \frac{-6 Q_\ell Q_t e^4}{c_W^2 s_W^2} \right) \left[ g_{At} g_{V\ell} \, \beta \cos\theta + 2 g_{A\ell} g_{Vt} \right] \sqrt{1 - \beta^2} \sin\theta &\\
&{} + \chi^2 \left( \frac{-6 e^4}{c_W^4 s_W^4} \right) \left[ 2 g_{A\ell} g_{V\ell} g_{Vt}^2  + g_{At} g_{Vt} (g_{A\ell}^2 + g_{V\ell}^2) \beta \cos\theta \right] \sqrt{1 - \beta^2} \sin\theta &
\end{flalign*}

\begin{flalign*}
A \, C_{kk} ={}& 3 Q_\ell^2 Q_t^2 e^4 \left[ \beta^2 + (2 - \beta^2)\cos^2\theta \right] &\\
&{} + \chi \frac{6 e^4 Q_\ell Q_t}{c_W^2 s_W^2} \left[ g_{V\ell} g_{Vt} \left( \beta^2 + (2 - \beta^2)\cos^2\theta \right) + 2 g_{A\ell} g_{At} \, \beta \cos\theta \right] &\\
&{} + \chi^2 \frac{3 e^4}{c_W^4 s_W^4} \left[ (g_{A\ell}^2 + g_{V\ell}^2) \left( g_{At}^2 \beta^2 (1 + \cos^2\theta) + g_{Vt}^2 \left( \beta^2 \sin^2\theta + 2\cos^2\theta \right) \right) + 8 g_{A\ell} g_{At} g_{V\ell} g_{Vt} \, \beta \cos\theta  \right] &
\end{flalign*}

\begin{flalign*}
A \, C_{kr} ={}& 6 Q_\ell^2 Q_t^2 e^4 \sqrt{1 - \beta^2} \sin\theta \cos\theta + \chi \frac{6 Q_\ell Q_t e^4}{c_W^2 s_W^2} \left[ 2 g_{V\ell} g_{Vt}  \cos\theta + \beta g_{A\ell} g_{At} \,  \right] \sqrt{1 - \beta^2} \sin\theta &\\
&{} + \chi^2 \frac{6 e^4}{c_W^4 s_W^4} \left[ g_{Vt}^2 (g_{A\ell}^2 + g_{V\ell}^2)  \cos\theta + 2 g_{A\ell} g_{At} g_{V\ell} g_{Vt} \, \beta  \right] \sqrt{1 - \beta^2} \sin\theta&
\end{flalign*}

\begin{flalign*}
A \, C_{rr} ={}& 3 Q_\ell^2 Q_t^2 e^4 (2 - \beta^2) \sin^2\theta + \chi \frac{6 Q_\ell Q_t e^4 g_{Vt} g_{V\ell}}{c_W^2 s_W^2} (2 - \beta^2) \sin^2\theta &\\
&{} + \chi^2 \frac{3 e^4}{c_W^4 s_W^4} \left[ 2 g_{Vt}^2 (g_{A\ell}^2 + g_{V\ell}^2) - \beta^2 (g_{At}^2 + g_{Vt}^2)(g_{A\ell}^2 + g_{V\ell}^2) \right] \sin^2\theta &
\end{flalign*}

\begin{flalign*}
A \, C_{nn} ={}& 3 Q_\ell^2 Q_t^2 e^4 \beta^2 (-\sin^2\theta) + \chi \frac{6 Q_\ell Q_t e^4 g_{V\ell} g_{Vt}}{c_W^2 s_W^2} \beta^2 (-\sin^2\theta) + \chi^2 \frac{3 e^4 \beta^2}{c_W^4 s_W^4} (g_{A\ell}^2 + g_{V\ell}^2)(g_{At}^2 - g_{Vt}^2) \sin^2\theta &
\end{flalign*}

Symmetry relates the remaining entries: $B^{-}_{k}=B^{+}_{k}$, $B^{-}_{r}=B^{+}_{r}$, $C_{rk}=C_{kr}$. All of $B^{+}_{n}$, $B^{-}_{n}$, $C_{kn}$, $C_{rn}$, $C_{nk}$, $C_{nr}$ vanish identically.

\paragraph{dipole interference.}

\begin{flalign*}
A ={}& 24\sqrt{2} \, Q_\ell^2 Q_t a_\gamma e^3 m_t + \chi \cdot 24\sqrt{2} \, Q_\ell e^3 m_t \left[ \frac{Q_t g_{V\ell} a_Z}{c_W s_W} + \frac{a_\gamma \left( g_{V\ell} g_{Vt} + \beta g_{A\ell} g_{At} \cos\theta \right)}{c_W^2 s_W^2} \right] &\\
&{}+ \chi^2 \cdot \frac{24\sqrt{2} \, e^3 a_Z m_t}{c_W^3 s_W^3} \left[ 2\beta g_{A\ell} g_{At} g_{V\ell} \cos\theta + \left(g_{A\ell}^2 + g_{V\ell}^2\right) g_{Vt} \right] &
\end{flalign*}

\begin{flalign*}
A \, B_k^+ ={}& \chi \cdot \left( -12\sqrt{2} \, Q_\ell e^3 m_t \right) \left[ \frac{2 Q_t g_{A\ell} a_Z \cos\theta}{c_W s_W} + \frac{\left( 2g_{A\ell} g_{Vt} \cos\theta + g_{At} g_{V\ell} \beta \left(1 + \cos^2\theta\right) \right) a_\gamma}{c_W^2 s_W^2}   \right] &\\
&{}+ \chi^2 \cdot \left( \frac{-12\sqrt{2} \, a_Z e^3 m_t}{c_W^3 s_W^3} \right) \left[ \beta g_{At} \left(g_{A\ell}^2 + g_{V\ell}^2\right) \left(1 + \cos^2\theta\right) + 4 g_{A\ell} g_{V\ell} g_{Vt} \cos\theta  \right] &
\end{flalign*}

\begin{flalign*}
A \, B_r^+ ={}& \chi \cdot \left( \frac{-12\sqrt{2} \, Q_\ell e^3 m_t}{ \sqrt{1-\beta^2}} \right) \left[ \frac{Q_t g_{A\ell} a_Z (2-\beta^2) \sin\theta}{c_W s_W}  + \frac{a_\gamma}{c_W^2 s_W^2} \left( g_{A\ell} g_{Vt} (2-\beta^2)\sin\theta + g_{At} g_{V\ell} \beta \sin\theta \cos\theta \right)  \right] &\\
&{}+ \chi^2 \cdot \left(\frac{-12\sqrt{2} \, a_Z e^3 m_t }{c_W^3 s_W^3 \sqrt{1-\beta^2}}\right) \left[ 2g_{A\ell} g_{V\ell} g_{Vt} (2-\beta^2)\sin\theta + \beta g_{At} \left(g_{A\ell}^2 + g_{V\ell}^2\right) \sin\theta \cos\theta   \right] &
\end{flalign*}

\begin{flalign*}
A \, B_n^+ ={}& \chi \cdot \left( \frac{12\sqrt{2} \, Q_\ell e^3 m_t \beta}{\sqrt{1-\beta^2}} \right) \left[ \frac{Q_t g_{A\ell} d_Z \sin\theta}{c_W s_W} + \frac{d_\gamma}{c_W^2 s_W^2} \left( g_{A\ell} g_{Vt} \sin\theta + \beta g_{At} g_{V\ell} \sin\theta \cos\theta \right)   \right] &\\
&{}+ \chi^2 \cdot \left( \frac{12\sqrt{2} \, d_Z e^3 m_t \beta }{c_W^3 s_W^3 \sqrt{1-\beta^2}} \right) \left[ 2g_{A\ell} g_{V\ell} g_{Vt} \sin\theta + \beta g_{At} \left(g_{A\ell}^2 + g_{V\ell}^2\right) \sin\theta \cos\theta  \right] &
\end{flalign*}

\begin{flalign*}
A \, C_{kk} ={}& 24\sqrt{2} \, Q_\ell^2 Q_t a_\gamma e^3 m_t \cos^2\theta &\\
&{}+ \chi \cdot 24\sqrt{2} \, e^3 Q_\ell m_t \left[ \frac{Q_t g_{V\ell} a_Z \cos^2\theta}{c_W s_W} + \frac{a_\gamma}{c_W^2 s_W^2} \left( g_{V\ell} g_{Vt} \cos^2\theta + g_{A\ell} g_{At} \beta \cos\theta \right)  \right] &\\
&{}+ \chi^2 \cdot \frac{24\sqrt{2} \, a_Z e^3 m_t}{c_W^3 s_W^3} \left[ g_{Vt} \left(g_{A\ell}^2 + g_{V\ell}^2\right) \cos^2\theta + 2g_{A\ell} g_{At} g_{V\ell} \beta \cos\theta \right] &
\end{flalign*}

\begin{flalign*}
A \, C_{kr} ={}& \frac{12\sqrt{2} \, Q_\ell^2 Q_t a_\gamma e^3 m_t \sqrt{1-\beta^2} \, (2-\beta^2)\sin\theta \cos\theta}{(1-\beta^2)} &\\
&{}+ \chi \cdot \frac{12\sqrt{2} \, Q_\ell e^3 m_t \sqrt{1-\beta^2}}{(1-\beta^2)} \left[ \frac{Q_t a_Z g_{V\ell} (2-\beta^2)\sin\theta \cos\theta}{c_W s_W} \right. & \\
&{}\left. + \frac{a_\gamma}{c_W^2 s_W^2} \left( g_{A\ell} g_{At} \beta \sin\theta + g_{V\ell} g_{Vt} (2-\beta^2)\sin\theta \cos\theta \right) \right] &\\
&{}+ \chi^2 \cdot \frac{12\sqrt{2} \, a_Z e^3 m_t \sqrt{1-\beta^2}}{c_W^3 s_W^3 (1-\beta^2)} \left[ g_{Vt} \left(g_{A\ell}^2 + g_{V\ell}^2\right) (2-\beta^2)\sin\theta \cos\theta + 2g_{A\ell} g_{At} g_{V\ell} \beta \sin\theta   \right] &
\end{flalign*}

\begin{flalign*}
A \, C_{rr} ={}& 24\sqrt{2} \, Q_\ell^2 Q_t a_\gamma e^3 m_t \sin^2\theta + \chi \cdot 24\sqrt{2} \, Q_\ell e^3 m_t \left[ \frac{Q_t g_{V\ell} a_Z}{c_W s_W} + \frac{g_{V\ell} g_{Vt} a_\gamma}{c_W^2 s_W^2} \right] \sin^2\theta &\\
&{}+ \chi^2 \cdot \frac{24\sqrt{2} \, a_Z e^3 g_{Vt} \left(g_{A\ell}^2 + g_{V\ell}^2\right) m_t \sin^2\theta}{c_W^3 s_W^3} &
\end{flalign*}

\begin{flalign*}
A \, C_{kn} ={}& \frac{12\sqrt{2} \, Q_\ell^2 Q_t e^3 m_t \beta \sqrt{1-\beta^2} \, d_\gamma \sin\theta \cos\theta}{(1-\beta^2)} &\\
&{}+ \chi \cdot \frac{12\sqrt{2} \, Q_\ell e^3 m_t \sqrt{1-\beta^2}}{(1-\beta^2)} \left[ \frac{\beta Q_t d_Z g_{V\ell} \sin\theta \cos\theta}{c_W s_W} + \frac{d_\gamma}{c_W^2 s_W^2} \left( g_{Vt} g_{V\ell} \beta \sin\theta \cos\theta + g_{A\ell} g_{At} \beta^2 \sin\theta \right)  \right] &\\
&{}+ \chi^2 \cdot \frac{12\sqrt{2} \, e^3 d_Z m_t \sqrt{1-\beta^2}}{(1-\beta^2) c_W^3 s_W^3} \left[ \beta g_{Vt} \left(g_{A\ell}^2 + g_{V\ell}^2\right) \sin\theta \cos\theta + 2\beta^2 g_{A\ell} g_{At} g_{V\ell} \sin\theta  \right] &
\end{flalign*}

\begin{flalign*}
A \, C_{rn} ={}& 12\sqrt{2} \, Q_\ell^2 Q_t e^3 d_\gamma m_t \beta \sin^2\theta + \chi \cdot 12\sqrt{2} \, Q_\ell e^3 m_t \beta \left[ \frac{Q_t g_{V\ell} d_Z}{c_W s_W} + \frac{g_{V\ell} g_{Vt} d_\gamma}{c_W^2 s_W^2} \right] \sin^2\theta &\\
&{}+ \chi^2 \cdot \frac{12\sqrt{2} \, e^3 g_{Vt} \left(g_{A\ell}^2 + g_{V\ell}^2\right) d_Z m_t \beta \sin^2\theta}{c_W^3 s_W^3} &
\end{flalign*}

Symmetry relates the remaining entries: $B^{-}_{k}=B^{+}_{k}$, $B^{-}_{r}=B^{+}_{r}$, $B^{-}_{n}=-B^{+}_{n}$, $C_{rk}=C_{kr}$, $C_{nk}=-C_{kn}$, $C_{nr}=-C_{rn}$. $C_{nn}$ vanishes identically.

\paragraph{dipole squared.}

\begin{flalign*}
A ={}& \frac{24 m_t^2 Q_\ell^2 e^2}{(1-\beta^2)} \left[ \left(\beta^2 \sin^2\theta - 2\beta^2 + 2\right) a_\gamma^2 + \beta^2 \sin^2\theta \, d_\gamma^2 \right] &\\
&{}+ \chi \cdot \frac{48 Q_\ell e^2 g_{V\ell} m_t^2}{(1-\beta^2) c_W s_W} \left[ \left(\beta^2 \sin^2\theta - 2\beta^2 + 2\right) a_\gamma a_Z + \beta^2 \sin^2\theta \, d_\gamma d_Z \right] &\\
&{}+ \chi^2 \cdot \frac{24 m_t^2 e^2 \left(g_{A\ell}^2 + g_{V\ell}^2\right)}{(1-\beta^2) c_W^2 s_W^2} \left[ \left(\beta^2 \sin^2\theta - 2\beta^2 + 2\right) a_Z^2 + \beta^2 d_Z^2 \sin^2\theta  \right] &
\end{flalign*}

\begin{flalign*}
A \, B_k^+ ={}& \chi \cdot \left( \frac{-96 m_t^2 Q_\ell a_\gamma a_Z e^2 g_{A\ell} \cos\theta}{c_W s_W} \right) + \chi^2 \cdot \left( \frac{-96 m_t^2 a_Z^2 e^2 g_{A\ell} g_{V\ell} \cos\theta}{c_W^2 s_W^2} \right) &
\end{flalign*}

\begin{flalign*}
A \, B_r^+ ={}& \chi \cdot \left( \frac{-96 Q_\ell a_\gamma a_Z e^2 g_{A\ell} m_t^2 \sin\theta}{c_W s_W \sqrt{1-\beta^2}} \right) + \chi^2 \cdot \left( \frac{-96 a_Z^2 e^2 g_{A\ell} g_{V\ell} m_t^2 \sin\theta}{c_W^2 s_W^2 \sqrt{1-\beta^2}} \right) &
\end{flalign*}

\begin{flalign*}
A \, B_n^+ ={}& \chi \cdot \frac{48 Q_\ell e^2 g_{A\ell} (a_\gamma d_Z + a_Z d_\gamma)}{c_W s_W} \frac{m_t^2 \beta \sin\theta}{\sqrt{1-\beta^2}} + \chi^2 \cdot \frac{96 a_Z d_Z e^2 g_{A\ell} g_{V\ell}}{c_W^2 s_W^2} \frac{m_t^2 \beta \sin\theta}{\sqrt{1-\beta^2}} &
\end{flalign*}

\begin{flalign*}
A \, C_{kk} ={}& \left( \frac{24 Q_\ell^2 e^2 m_t^2}{(1-\beta^2)} \right) \left[ \left(2\cos^2\theta + \beta^2 (\sin^2\theta - 2)\right) a_\gamma^2 - \beta^2 \sin^2\theta \, d_\gamma^2 \right] &\\
&{}+ \chi \cdot \left( \frac{48 Q_\ell e^2 m_t^2 g_{V\ell}}{c_W s_W (1-\beta^2)} \right) \left[ \left(2\cos^2\theta + \beta^2 (\sin^2\theta - 2)\right) a_\gamma a_Z - \beta^2 \sin^2\theta \, d_\gamma d_Z \right] &\\
&{}+ \chi^2 \cdot \left( \frac{24 e^2 (g_{A\ell}^2 + g_{V\ell}^2) m_t^2}{c_W^2 s_W^2 (1-\beta^2)} \right) \left[ \left(2\cos^2\theta + \beta^2 (\sin^2\theta - 2)\right) a_Z^2 - \beta^2 \sin^2\theta \, d_Z^2 \right] &
\end{flalign*}

\begin{flalign*}
A \, C_{kr} ={}& \left(24 Q_\ell^2 a_\gamma^2 e^2\right) \frac{m_t^2 \sin 2\theta}{\sqrt{1-\beta^2}} + \chi \cdot \left( \frac{48 Q_\ell a_\gamma a_Z e^2 g_{V\ell}}{c_W s_W} \right) \frac{m_t^2 \sin 2\theta}{\sqrt{1-\beta^2} } + \chi^2 \cdot \left( \frac{24 a_Z^2 e^2 (g_{A\ell}^2 + g_{V\ell}^2)}{c_W^2 s_W^2} \right) \frac{m_t^2 \sin 2\theta}{\sqrt{1-\beta^2} } &
\end{flalign*}

\begin{flalign*}
A \, C_{rr} ={}& \left( \frac{24 Q_\ell^2 e^2 m_t^2}{1-\beta^2} \right) \left[ (2-\beta^2)\sin^2\theta \, a_\gamma^2 - \beta^2 \sin^2\theta \, d_\gamma^2 \right] &\\
&{}+ \chi \cdot \left( \frac{48 Q_\ell e^2 g_{V\ell} m_t^2}{c_W s_W (1-\beta^2)} \right) \left[ (2-\beta^2)\sin^2\theta \, a_\gamma a_Z - \beta^2 \sin^2\theta \, d_\gamma d_Z \right] &\\
&{}+ \chi^2 \cdot \left( \frac{24 e^2 (g_{A\ell}^2 + g_{V\ell}^2) m_t^2}{c_W^2 s_W^2 (1-\beta^2)} \right) \left[ (2-\beta^2)\sin^2\theta \, a_Z^2 - \beta^2 \sin^2\theta \, d_Z^2 \right] &
\end{flalign*}

\begin{flalign*}
A \, C_{kn} ={}& 24 Q_\ell^2 a_\gamma d_\gamma e^2 \, \frac{m_t^2 \beta \sin 2\theta}{\sqrt{1-\beta^2} } + \chi \cdot \left( \frac{24 Q_\ell e^2 g_{V\ell} (a_\gamma d_Z + a_Z d_\gamma)}{c_W s_W} \right) \frac{m_t^2 \beta \sin 2\theta}{\sqrt{1-\beta^2} } &\\
&{}+ \chi^2 \cdot \left( \frac{24 a_Z d_Z e^2 (g_{A\ell}^2 + g_{V\ell}^2)}{c_W^2 s_W^2} \right) \frac{m_t^2 \beta \sin 2\theta}{\sqrt{1-\beta^2}} &
\end{flalign*}

\begin{flalign*}
A \, C_{rn} ={}& 48 Q_\ell^2 a_\gamma d_\gamma e^2 \, \frac{m_t^2 \beta \sin^2\theta}{(1-\beta^2)} + \chi \cdot \left( \frac{48 Q_\ell e^2 g_{V\ell} (a_\gamma d_Z + a_Z d_\gamma)}{c_W s_W} \right) \frac{m_t^2 \beta \sin^2\theta}{(1-\beta^2)} &\\
&{}+ \chi^2 \cdot \left( \frac{48 a_Z d_Z e^2 (g_{A\ell}^2 + g_{V\ell}^2)}{c_W^2 s_W^2} \right) \frac{m_t^2 \beta \sin^2\theta}{(1-\beta^2)} &
\end{flalign*}

\begin{flalign*}
A \, C_{nn} ={}& 24 Q_\ell^2 e^2 \left(a_\gamma^2 - d_\gamma^2\right) \frac{m_t^2 \beta^2 \sin^2\theta}{1-\beta^2} + \chi \cdot \left( \frac{48 Q_\ell e^2 g_{V\ell} (a_\gamma a_Z - d_\gamma d_Z)}{c_W s_W} \right) \frac{m_t^2 \beta^2 \sin^2\theta}{1-\beta^2} &\\
&{}+ \chi^2 \cdot \left( \frac{24 e^2 (g_{A\ell}^2 + g_{V\ell}^2)(a_Z^2 - d_Z^2)}{c_W^2 s_W^2} \right) \frac{m_t^2 \beta^2 \sin^2\theta}{1-\beta^2} &
\end{flalign*}

Symmetry relates the remaining entries: $B^{-}_{k}=B^{+}_{k}$, $B^{-}_{r}=B^{+}_{r}$, $B^{-}_{n}=-B^{+}_{n}$, $C_{rk}=C_{kr}$, $C_{nk}=-C_{kn}$, $C_{nr}=-C_{rn}$.

\subsection{$q\bar q\to\ttbar$}

\paragraph{SM.}

\begin{flalign*}
A &= \frac{2}{9} g_s^4 \left( 2 - \beta^2 \sin^2\theta \right) &
\end{flalign*}

\begin{flalign*}
A \, C_{kk} &= \frac{2}{9} g_s^4 \left[ \beta^2 + (2 - \beta^2)\cos^2\theta \right] &
\end{flalign*}

\begin{flalign*}
A \, C_{kr} &= \frac{2}{9} g_s^4 \sqrt{1 - \beta^2} \sin(2\theta) &
\end{flalign*}

\begin{flalign*}
A \, C_{rr} &= \frac{2}{9} g_s^4 (2 - \beta^2) \sin^2\theta &
\end{flalign*}

\begin{flalign*}
A \, C_{nn} &= -\frac{2}{9} g_s^4 \beta^2 \sin^2\theta &
\end{flalign*}

Symmetry relates the remaining entries: $C_{rk}=C_{kr}$. All of $B^{+}_{k}$, $B^{+}_{r}$, $B^{+}_{n}$, $B^{-}_{k}$, $B^{-}_{r}$, $B^{-}_{n}$, $C_{kn}$, $C_{rn}$, $C_{nk}$, $C_{nr}$ vanish identically.

\paragraph{dipole interference.}

\begin{flalign*}
A &= \frac{16\sqrt{2}\, a_g g_s^4 m_t}{9} &
\end{flalign*}

\begin{flalign*}
A \, C_{kk} &= \frac{16\sqrt{2}\, g_s^4 a_g m_t \cos^2\theta}{9} &
\end{flalign*}

\begin{flalign*}
A \, C_{kr} &= \frac{4\sqrt{2}\, g_s^4 a_g m_t \sqrt{1 - \beta^2}}{9} \left( \frac{2 - \beta^2}{1 - \beta^2} \right) \sin(2\theta) &
\end{flalign*}

\begin{flalign*}
A \, C_{rr} &= \frac{16\sqrt{2}\, g_s^4 a_g m_t \sin^2\theta}{9} &
\end{flalign*}

\begin{flalign*}
A \, C_{kn} &= \frac{8\sqrt{2}\, g_s^4 d_g m_t \beta}{\sqrt{1 - \beta^2}} \frac{\sin(2\theta)}{18} &
\end{flalign*}

\begin{flalign*}
A \, C_{rn} &= \frac{8\sqrt{2}\, g_s^4 d_g m_t \beta \sin^2\theta}{9} &
\end{flalign*}

Symmetry relates the remaining entries: $C_{rk}=C_{kr}$, $C_{nk}=-C_{kn}$, $C_{nr}=-C_{rn}$. All of $B^{+}_{k}$, $B^{+}_{r}$, $B^{+}_{n}$, $B^{-}_{k}$, $B^{-}_{r}$, $B^{-}_{n}$, $C_{nn}$ vanish identically.

\paragraph{dipole squared.}

\begin{flalign*}
A &= \frac{16 g_s^4 m_t^2}{9(1-\beta^2)} \left[ \left( 2 - \beta^2 (1 + \cos^2\theta) \right) a_g^2 + d_g^2 \beta^2 \sin^2\theta \right] &
\end{flalign*}

\begin{flalign*}
A \, C_{kk} &= \frac{16 g_s^4 m_t^2}{9(1-\beta^2)} \left[ \left( (2 - \beta^2)\cos^2\theta - \beta^2 \right) a_g^2 - \beta^2 \sin^2\theta \, d_g^2 \right] &
\end{flalign*}

\begin{flalign*}
A \, C_{kr} &= \frac{16 g_s^4 m_t^2}{9(1-\beta^2)} \left( \sqrt{1 - \beta^2} \sin 2\theta \right) a_g^2 &
\end{flalign*}

\begin{flalign*}
A \, C_{rr} &= \frac{16 g_s^4 m_t^2}{9(1-\beta^2)} \left[ (2 - \beta^2) \sin^2\theta \, a_g^2 - \beta^2 \sin^2\theta \, d_g^2 \right] &
\end{flalign*}

\begin{flalign*}
A \, C_{kn} &= \frac{16 g_s^4 m_t^2}{9(1-\beta^2)} \left( \beta \sqrt{1 - \beta^2} \sin 2\theta \right) a_g d_g &
\end{flalign*}

\begin{flalign*}
A \, C_{rn} &= \frac{16 g_s^4 m_t^2}{9(1-\beta^2)} \left( 2\beta \sin^2\theta \right) a_g d_g &
\end{flalign*}

\begin{flalign*}
A \, C_{nn} &= \frac{16 g_s^4 m_t^2}{9(1-\beta^2)} (\beta^2 \sin^2\theta) (a_g^2 - d_g^2) &
\end{flalign*}

Symmetry relates the remaining entries: $C_{rk}=C_{kr}$, $C_{nk}=-C_{kn}$, $C_{nr}=-C_{rn}$. All of $B^{+}_{k}$, $B^{+}_{r}$, $B^{+}_{n}$, $B^{-}_{k}$, $B^{-}_{r}$, $B^{-}_{n}$ vanish identically.

\subsection{$gg\to\ttbar$}

\paragraph{SM.}
\begin{flalign*}
A ={} &\frac{g_s^4 \left(7 + 9\beta^2 \cos^2\theta\right)}{12 (1 - \beta^2 \cos^2\theta)^2} \Big[ 1 + 2\beta^2 \sin^2\theta - \beta^4 (1 + \sin^4\theta) \Big] &&
 \end{flalign*}

\begin{flalign*}
A\,C_{kk} ={} & -\frac{g_s^4 \left(7 + 9\beta^2 \cos^2\theta\right)}{12 (1 - \beta^2 \cos^2\theta)^2} \left[ 1 - \beta^2 \frac{\sin^2 2\theta}{2} - \beta^4 (1 + \sin^4\theta) \right] &&
\end{flalign*}
\begin{flalign*}
A\,C_{rk} ={} & \frac{g_s^4 \left(7 + 9\beta^2 \cos^2\theta\right)}{12 (1 - \beta^2 \cos^2\theta)^2} \left[ \beta^2 \sqrt{1-\beta^2} \,  \sin 2\theta \sin^2\theta  \right] &&
\end{flalign*}

\begin{flalign*}
A\,C_{rr} ={} & -\frac{g_s^4 \left(7 + 9\beta^2 \cos^2\theta\right)}{12 (1 - \beta^2 \cos^2\theta)^2} \left[ 1 - \beta^2 (2 - \beta^2) (1 + \sin^4\theta) \right] &&
\end{flalign*}
\begin{flalign*}
A\,C_{nn} ={} & -\frac{g_s^4 \left(7 + 9\beta^2 \cos^2\theta\right)}{12 (1 - \beta^2 \cos^2\theta)^2} \left[ 1 - 2\beta^2 + \beta^4 (1 + \sin^4\theta) \right] &&
\end{flalign*}

Symmetry relates the remaining entries: $C_{rk}=C_{kr}$. All of $B^{+}_{k}$, $B^{+}_{r}$, $B^{+}_{n}$, $B^{-}_{k}$, $B^{-}_{r}$, $B^{-}_{n}$, $C_{kn}$, $C_{rn}$, $C_{nk}$, $C_{nr}$ vanish identically.

\paragraph{dipole interference.}

\begin{flalign*}
A ={} & -\frac{2\sqrt{2} \, a_g g_s^4 m_t}{3(\beta^2 \cos^2\theta - 1)} \left[ 7 + 9\beta^2 \cos^2\theta \right] &&
\end{flalign*}

\begin{flalign*}
A\,C_{kk} ={} & -\frac{2\sqrt{2}}{3} \frac{a_g g_s^4 m_t}{(\beta^2 \cos^2\theta - 1)^2} \left(7 + 9\beta^2 \cos^2\theta\right) \left(1 - \beta^2 \left(1 + \cos^2\theta - \cos^4\theta\right)\right) &&
\end{flalign*}

\begin{flalign*}
A\,C_{kr} ={} & \frac{\sqrt{2} \, a_g g_s^4}{3 \left(1 - \beta^2 \cos^2\theta\right)^2} \frac{m_t \sqrt{1 - \beta^2}}{(1 - \beta^2)} \cot\theta\\ &\Big[ - 9 \beta^{4} (\beta^{2} - 2) \, \cos^6\theta + \beta^{2} (\beta^{2} - 2) (18 \beta^{2} - 7) \, \cos^4\theta - \beta^{2} (9 \beta^{4} - 16 \beta^{2} + 12) \, \cos^2\theta + \beta^{2} (9 \beta^{2} - 2) \Big] &&
\end{flalign*}

\begin{flalign*}
A\,C_{rr} ={} & \frac{-2\sqrt{2} \, a_g g_s^4 m_t}{3 \left(1 - \beta^2 \cos^2\theta\right)^2} \left[ 7\left(1 - \beta^2\right) + \sin^2\theta \cos^2\theta \, \beta^2 \left(7 - 9\beta^2 \sin^2\theta\right)  \right] &&
\end{flalign*}

\begin{flalign*}
A\,C_{kn} ={} & \frac{\sqrt{2} \, d_g g_s^4}{6 \left(1 - \beta^2 \cos^2\theta\right)^2} \frac{m_t \sqrt{1 - \beta^2}}{(1 - \beta^2)} \Big[ - 18 \beta^{5} \, \sin\theta \, \cos^5\theta  && \\
& + 2 \beta^{3} (9 \beta^{2} - 7) \, \sin\theta \, \cos^3\theta + 2 \beta (23 \beta^{2} - 16) \, \sin\theta \, \cos\theta \Big] &&
\end{flalign*}

\begin{flalign*}
A\,C_{rn} ={} & \frac{\sqrt{2} \, d_g g_s^4 m_t \beta}{3 \left(1 - \beta^2 \cos^2\theta\right)^2} \Big[ 9 \beta^{4} \, \cos^6\theta - \beta^{2} (18 \beta^{2} - 7) \, \cos^4\theta + (18 \beta^{4} - 25 \beta^{2} + 16) \, \cos^2\theta + 7 (2 \beta^{2} - 3) \Big] &&
\end{flalign*}

\begin{flalign*}
A\,C_{nn} ={} & \frac{-14\sqrt{2} \, a_g g_s^4 m_t}{3 \left(1 - \beta^2 \cos^2\theta\right)} &&
\end{flalign*}

Symmetry relates the remaining entries: $C_{rk}=C_{kr}$, $C_{nk}=-C_{kn}$, $C_{nr}=-C_{rn}$. All of $B^{+}_{k}$, $B^{+}_{r}$, $B^{+}_{n}$, $B^{-}_{k}$, $B^{-}_{r}$, $B^{-}_{n}$ vanish identically.

\paragraph{dipole squared.}

\begin{flalign*}
A ={} & \frac{2 m_t^2 g_s^4}{3 \left(\beta ^2 \cos ^2\theta - 1\right)^2 \left(1 - \beta^2\right)} \Big[ a_g^2 \Big( 9 \beta^{6} \, \cos^6\theta + 3 \beta^{4} \, \cos^4\theta  && \\
& + \beta^{2} (4 \beta - 7) (4 \beta + 7) \, \cos^2\theta + (- 16 \beta^{2} + 37) \Big) + d_g^2 \Big( 9 \beta^{6} \, \cos^6\theta - 3 \beta^{4} (6 \beta^{2} - 7) \, \cos^4\theta  && \\
& + \beta^{2} (18 \beta^{4} - 16 \beta^{2} - 35) \, \cos^2\theta + (14 \beta^{4} - 30 \beta^{2} + 37) \Big) \Big] &&
\end{flalign*}

\begin{flalign*}
A\,C_{kk} ={} & \frac{2 m_t^2 g_s^4 \beta^2}{3 \left(\beta ^2 \cos ^2\theta - 1\right)^2 \left(1 - \beta^2\right)} \Big[ a_g^2 \Big( 9 \beta^{2} (\beta^{2} - 2) \, \cos^6\theta + (25 \beta^{2} - 28) \, \cos^4\theta  && \\
& - \frac{3 (6 \beta^{6} + 4 \beta^{4} - 27 \beta^{2} + 6)}{\beta^{2}} \, \cos^2\theta - \frac{14 \beta^{4} - 12 \beta^{2} + 19}{\beta^{2}} \Big)  && \\
& + d_g^2 \Big( - 9 \beta^{4} \, \cos^6\theta + (18 \beta^{4} - 7 \beta^{2} - 14) \, \cos^4\theta && \\
& - \frac{(\beta^{2} + 2) (18 \beta^{4} - 38 \beta^{2} + 9)}{\beta^{2}} \, \cos^2\theta - \frac{14 \beta^{4} - 12 \beta^{2} + 19}{\beta^{2}} \Big) \Big] &&
\end{flalign*}

\begin{flalign*}
A\,C_{kr} ={} & \frac{m_t^2 g_s^4 \cot\theta \sqrt{1-\beta^2}}{24 \left(\beta^2 \cos^2\theta - 1\right)^2 \left(1 - \beta^2\right)} \Big[ a_g^2 \Big( 288 \beta^{4} \, \cos^6\theta - 64 \beta^{2} (9 \beta^{2} - 7) \, \cos^4\theta  && \\
& + 32 (9 \beta^{4} - 37 \beta^{2} + 9) \, \cos^2\theta + 32 (23 \beta^{2} - 9) \Big) && \\
& + d_g^2 \Big( 224 \beta^{2} \, \cos^4\theta - 32 (23 \beta^{2} - 9) \, \cos^2\theta + 32 (4 \beta - 3) (4 \beta + 3) \Big) \Big] &&
\end{flalign*}
\begin{flalign*}
A\,C_{rr} ={} & \frac{2 m_t^2 g_s^4}{3 \left(\beta ^2 \cos ^2\theta - 1\right)^2 \left(1 - \beta^2\right)} \Big[ a_g^2 \Big( - 9 \beta^{4} (\beta^{2} - 2) \, \cos^6\theta + \beta^{2} (18 \beta^{4} - 57 \beta^{2} + 28) \, \cos^4\theta && \\
& + (- 18 \beta^{6} + 52 \beta^{4} - 57 \beta^{2} + 18) \, \cos^2\theta + (- 14 \beta^{4} + 58 \beta^{2} - 37) \Big)  && \\
& + d_g^2 \Big( - 9 \beta^{6} \, \cos^6\theta+ \beta^{2} (18 \beta^{4} - 21 \beta^{2} + 14) \, \cos^4\theta && \\
& + (- 18 \beta^{6} + 34 \beta^{4} - 29 \beta^{2} + 18) \, \cos^2\theta + (- 14 \beta^{4} + 44 \beta^{2} - 37) \Big) \Big] &&
\end{flalign*}
\begin{flalign*}
A\,C_{kn} ={} & \frac{4 m_t^2 a_g d_g g_s^4 \beta^3\sqrt{1-\beta^2}}{3 \left(\beta ^2 \cos ^2\theta - 1\right)^2 \left(1 - \beta^2\right)} \Big[ - 9 \beta^{2} \, \sin\theta \, \cos^5\theta + (9 \beta^{2} - 7) \, \sin\theta \, \cos^3\theta + 7 \, \sin\theta \, \cos\theta \Big] &&
\end{flalign*}
\begin{flalign*}
A\,C_{rn} ={} & \frac{4 m_t^2 a_g d_g g_s^4 \beta^3}{3 \left(\beta ^2 \cos ^2\theta - 1\right)^2 \left(1 - \beta^2\right)} \Big[ 9 \beta^{2} \, \cos^6\theta + (- 18 \beta^{2} + 7) \, \cos^4\theta + (9 \beta^{2} - 14) \, \cos^2\theta + 7 \Big] &&
\end{flalign*}
\begin{flalign*}
A\,C_{nn} ={} & \frac{2 m_t^2 g_s^4}{3 \left(\beta ^2 \cos ^2\theta - 1\right)^2 \left(1 - \beta^2\right)} \Big[ a_g^2 \Big(9 \beta^6 \cos^4\theta \left(\cos^2\theta - 2\right) && \\
& + \beta^4 \cos^2\theta \left(7 \cos^2\theta - 8\right) + \beta^2 \left(3 \cos^2\theta + 26\right) - 19\Big) && \\
& + d_g^2 \Big(-9 \beta^6 \cos^2\theta \left(\cos^4\theta - 2 \cos^2\theta + 2\right) + \beta^4 \left(-7 \cos^4\theta + 20 \cos^2\theta - 14\right) && \\
& + \beta^2 \left(3 \cos^2\theta + 26\right) - 19 \Big) \Big] &&
\end{flalign*}

Symmetry relates the remaining entries: $C_{rk}=C_{kr}$, $C_{nk}=-C_{kn}$, $C_{nr}=-C_{rn}$. All of $B^{+}_{k}$, $B^{+}_{r}$, $B^{+}_{n}$, $B^{-}_{k}$, $B^{-}_{r}$, $B^{-}_{n}$ vanish identically.

}

\bibliographystyle{JHEP}
\bibliography{refs}{}

\providecommand{\href}[2]{#2}\begingroup\raggedright\begin{thebibliography}{10}

\bibitem{Sakharov:1967dj}
A.D.~Sakharov, \emph{{Violation of CP Invariance, C asymmetry, and baryon
  asymmetry of the universe}},
  \href{https://doi.org/10.1070/PU1991v034n05ABEH002497}{\emph{Pisma Zh. Eksp.
  Teor. Fiz.} {\bfseries 5} (1967) 32}.

\bibitem{Farrar:1993hn}
G.R.~Farrar and M.E.~Shaposhnikov, \emph{{Baryon asymmetry of the universe in
  the standard electroweak theory}},
  \href{https://doi.org/10.1103/PhysRevD.50.774}{\emph{Phys. Rev. D} {\bfseries
  50} (1994) 774} [\href{https://arxiv.org/abs/hep-ph/9305275}{{\ttfamily
  hep-ph/9305275}}].

\bibitem{Atwood:2000tu}
D.~Atwood, S.~Bar-Shalom, G.~Eilam and A.~Soni, \emph{{CP violation in top
  physics}}, \href{https://doi.org/10.1016/S0370-1573(00)00112-5}{\emph{Phys.
  Rept.} {\bfseries 347} (2001) 1}
  [\href{https://arxiv.org/abs/hep-ph/0006032}{{\ttfamily hep-ph/0006032}}].

\bibitem{Kamenik:2011dk}
J.F.~Kamenik, M.~Papucci and A.~Weiler, \emph{{Constraining the dipole moments
  of the top quark}},
  \href{https://doi.org/10.1103/PhysRevD.85.071501}{\emph{Phys. Rev. D}
  {\bfseries 85} (2012) 071501}
  [\href{https://arxiv.org/abs/1107.3143}{{\ttfamily 1107.3143}}].

\bibitem{Czarnecki:1997bu}
A.~Czarnecki and B.~Krause, \emph{{Neutron electric dipole moment in the
  standard model: Valence quark contributions}},
  \href{https://doi.org/10.1103/PhysRevLett.78.4339}{\emph{Phys. Rev. Lett.}
  {\bfseries 78} (1997) 4339}
  [\href{https://arxiv.org/abs/hep-ph/9704355}{{\ttfamily hep-ph/9704355}}].

\bibitem{Gupta:2009wu}
S.K.~Gupta, A.S.~Mete and G.~Valencia, \emph{{CP violating anomalous top-quark
  couplings at the LHC}},
  \href{https://doi.org/10.1103/PhysRevD.80.034013}{\emph{Phys. Rev. D}
  {\bfseries 80} (2009) 034013}
  [\href{https://arxiv.org/abs/0905.1074}{{\ttfamily 0905.1074}}].

\bibitem{ATLAS:2023fsd}
{\scshape ATLAS} collaboration, \emph{{Observation of quantum entanglement with
  top quarks at the ATLAS detector}},
  \href{https://doi.org/10.1038/s41586-024-07824-z}{\emph{Nature} {\bfseries
  633} (2024) 542} [\href{https://arxiv.org/abs/2311.07288}{{\ttfamily
  2311.07288}}].

\bibitem{CMS:2024pts}
{\scshape CMS} collaboration, \emph{{Observation of quantum entanglement in top
  quark pair production in proton--proton collisions at $\sqrt{s} = 13$ TeV}},
  \href{https://doi.org/10.1088/1361-6633/ad7e4d}{\emph{Rept. Prog. Phys.}
  {\bfseries 87} (2024) 117801}
  [\href{https://arxiv.org/abs/2406.03976}{{\ttfamily 2406.03976}}].

\bibitem{Kane:1991bg}
G.L.~Kane, G.A.~Ladinsky and C.P.~Yuan, \emph{{Using the Top Quark for Testing
  Standard Model Polarization and CP Predictions}},
  \href{https://doi.org/10.1103/PhysRevD.45.124}{\emph{Phys. Rev. D} {\bfseries
  45} (1992) 124}.

\bibitem{Bernreuther:1993hq}
W.~Bernreuther and A.~Brandenburg, \emph{{Tracing CP violation in the
  production of top quark pairs by multiple TeV proton proton collisions}},
  \href{https://doi.org/10.1103/PhysRevD.49.4481}{\emph{Phys. Rev. D}
  {\bfseries 49} (1994) 4481}
  [\href{https://arxiv.org/abs/hep-ph/9312210}{{\ttfamily hep-ph/9312210}}].

\bibitem{Mahlon:1995zn}
G.~Mahlon and S.J.~Parke, \emph{{Angular correlations in top quark pair
  production and decay at hadron colliders}},
  \href{https://doi.org/10.1103/PhysRevD.53.4886}{\emph{Phys. Rev. D}
  {\bfseries 53} (1996) 4886}
  [\href{https://arxiv.org/abs/hep-ph/9512264}{{\ttfamily hep-ph/9512264}}].

\bibitem{Bernreuther:2015yna}
W.~Bernreuther, D.~Heisler and Z.-G.~Si, \emph{{A set of top quark spin
  correlation and polarization observables for the LHC: Standard Model
  predictions and new physics contributions}},
  \href{https://doi.org/10.1007/JHEP12(2015)026}{\emph{JHEP} {\bfseries 12}
  (2015) 026} [\href{https://arxiv.org/abs/1508.05271}{{\ttfamily
  1508.05271}}].

\bibitem{ATLAS:2012ao}
{\scshape ATLAS} collaboration, \emph{{Observation of spin correlation in $t
  \bar{t}$ events from pp collisions at sqrt(s) = 7 TeV using the ATLAS
  detector}}, \href{https://doi.org/10.1103/PhysRevLett.108.212001}{\emph{Phys.
  Rev. Lett.} {\bfseries 108} (2012) 212001}
  [\href{https://arxiv.org/abs/1203.4081}{{\ttfamily 1203.4081}}].

\bibitem{ATLAS:2019zrq}
{\scshape ATLAS} collaboration, \emph{{Measurements of top-quark pair spin
  correlations in the $e\mu$ channel at $\sqrt{s} = 13$ TeV using $pp$
  collisions in the ATLAS detector}},
  \href{https://doi.org/10.1140/epjc/s10052-020-8181-6}{\emph{Eur. Phys. J. C}
  {\bfseries 80} (2020) 754}
  [\href{https://arxiv.org/abs/1903.07570}{{\ttfamily 1903.07570}}].

\bibitem{CMS:2019nrx}
{\scshape CMS} collaboration, \emph{{Measurement of the top quark polarization
  and $\mathrm{t\bar{t}}$ spin correlations using dilepton final states in
  proton-proton collisions at $\sqrt{s} =$ 13 TeV}},
  \href{https://doi.org/10.1103/PhysRevD.100.072002}{\emph{Phys. Rev. D}
  {\bfseries 100} (2019) 072002}
  [\href{https://arxiv.org/abs/1907.03729}{{\ttfamily 1907.03729}}].

\bibitem{Czarnecki:1990pe}
A.~Czarnecki, M.~Jezabek and J.H.~Kuhn, \emph{{Lepton Spectra From Decays of
  Polarized Top Quarks}},
  \href{https://doi.org/10.1016/0550-3213(91)90082-9}{\emph{Nucl. Phys. B}
  {\bfseries 351} (1991) 70}.

\bibitem{Afik:2020onf}
Y.~Afik and J.R.M.~de~Nova, \emph{{Entanglement and quantum tomography with top
  quarks at the LHC}},
  \href{https://doi.org/10.1140/epjp/s13360-021-01902-1}{\emph{Eur. Phys. J.
  Plus} {\bfseries 136} (2021) 907}
  [\href{https://arxiv.org/abs/2003.02280}{{\ttfamily 2003.02280}}].

\bibitem{Rahaman:2021fcz}
R.~Rahaman and R.K.~Singh, \emph{{Breaking down the entire spectrum of spin
  correlations of a pair of particles involving fermions and gauge bosons}},
  \href{https://doi.org/10.1016/j.nuclphysb.2022.115984}{\emph{Nucl. Phys. B}
  {\bfseries 984} (2022) 115984}
  [\href{https://arxiv.org/abs/2109.09345}{{\ttfamily 2109.09345}}].

\bibitem{Afik:2022kwm}
Y.~Afik and J.R.M.~de~Nova, \emph{{Quantum information with top quarks in
  QCD}}, \href{https://doi.org/10.22331/q-2022-09-29-820}{\emph{Quantum}
  {\bfseries 6} (2022) 820} [\href{https://arxiv.org/abs/2203.05582}{{\ttfamily
  2203.05582}}].

\bibitem{Fabbrichesi:2021npl}
M.~Fabbrichesi, R.~Floreanini and G.~Panizzo, \emph{{Testing Bell Inequalities
  at the LHC with Top-Quark Pairs}},
  \href{https://doi.org/10.1103/PhysRevLett.127.161801}{\emph{Phys. Rev. Lett.}
  {\bfseries 127} (2021) 161801}
  [\href{https://arxiv.org/abs/2102.11883}{{\ttfamily 2102.11883}}].

\bibitem{Severi:2021cnj}
C.~Severi, C.D.E.~Boschi, F.~Maltoni and M.~Sioli, \emph{{Quantum tops at the
  LHC: from entanglement to Bell inequalities}},
  \href{https://doi.org/10.1140/epjc/s10052-022-10245-9}{\emph{Eur. Phys. J. C}
  {\bfseries 82} (2022) 285}
  [\href{https://arxiv.org/abs/2110.10112}{{\ttfamily 2110.10112}}].

\bibitem{Aguilar-Saavedra:2022uye}
J.A.~Aguilar-Saavedra and J.A.~Casas, \emph{{Improved tests of entanglement and
  Bell inequalities with LHC tops}},
  \href{https://doi.org/10.1140/epjc/s10052-022-10630-4}{\emph{Eur. Phys. J. C}
  {\bfseries 82} (2022) 666}
  [\href{https://arxiv.org/abs/2205.00542}{{\ttfamily 2205.00542}}].

\bibitem{Afik:2022dgh}
Y.~Afik and J.R.M.~de~Nova, \emph{{Quantum Discord and Steering in Top Quarks
  at the LHC}},
  \href{https://doi.org/10.1103/PhysRevLett.130.221801}{\emph{Phys. Rev. Lett.}
  {\bfseries 130} (2023) 221801}
  [\href{https://arxiv.org/abs/2209.03969}{{\ttfamily 2209.03969}}].

\bibitem{Han:2024ugl}
T.~Han, M.~Low, N.~McGinnis and S.~Su, \emph{{Measuring quantum discord at the
  LHC}}, \href{https://doi.org/10.1007/JHEP05(2025)081}{\emph{JHEP} {\bfseries
  05} (2025) 081} [\href{https://arxiv.org/abs/2412.21158}{{\ttfamily
  2412.21158}}].

\bibitem{Barr:2024djo}
A.J.~Barr, M.~Fabbrichesi, R.~Floreanini, E.~Gabrielli and L.~Marzola,
  \emph{{Quantum entanglement and Bell inequality violation at colliders}},
  \href{https://doi.org/10.1016/j.ppnp.2024.104134}{\emph{Prog. Part. Nucl.
  Phys.} {\bfseries 139} (2024) 104134}
  [\href{https://arxiv.org/abs/2402.07972}{{\ttfamily 2402.07972}}].

\bibitem{Aoude:2022imd}
R.~Aoude, E.~Madge, F.~Maltoni and L.~Mantani, \emph{{Quantum SMEFT tomography:
  Top quark pair production at the LHC}},
  \href{https://doi.org/10.1103/PhysRevD.106.055007}{\emph{Phys. Rev. D}
  {\bfseries 106} (2022) 055007}
  [\href{https://arxiv.org/abs/2203.05619}{{\ttfamily 2203.05619}}].

\bibitem{Severi:2022qjy}
C.~Severi and E.~Vryonidou, \emph{{Quantum entanglement and top spin
  correlations in SMEFT at higher orders}},
  \href{https://doi.org/10.1007/JHEP01(2023)148}{\emph{JHEP} {\bfseries 01}
  (2023) 148} [\href{https://arxiv.org/abs/2210.09330}{{\ttfamily
  2210.09330}}].

\bibitem{Maltoni:2024tul}
F.~Maltoni, C.~Severi, S.~Tentori and E.~Vryonidou, \emph{{Quantum detection of
  new physics in top-quark pair production at the LHC}},
  \href{https://doi.org/10.1007/JHEP03(2024)099}{\emph{JHEP} {\bfseries 03}
  (2024) 099} [\href{https://arxiv.org/abs/2401.08751}{{\ttfamily
  2401.08751}}].

\bibitem{Maltoni:2024csn}
F.~Maltoni, C.~Severi, S.~Tentori and E.~Vryonidou, \emph{{Quantum tops at
  circular lepton colliders}},
  \href{https://doi.org/10.1007/JHEP09(2024)001}{\emph{JHEP} {\bfseries 09}
  (2024) 001} [\href{https://arxiv.org/abs/2404.08049}{{\ttfamily
  2404.08049}}].

\bibitem{Cao:2025xnp}
H.~Cao and F.~Petriello, \emph{{Single-spin measurements and heavy new physics
  in the e+e- to ttbar process at an FCC-ee}},
  \href{https://doi.org/10.1103/794y-gp3r}{\emph{Phys. Rev. D} {\bfseries 113}
  (2026) 035033} [\href{https://arxiv.org/abs/2511.01994}{{\ttfamily
  2511.01994}}].

\bibitem{Bernreuther:1992be}
W.~Bernreuther, O.~Nachtmann, P.~Overmann and T.~Schr{\"o}der, \emph{{Angular
  correlations and distributions for searches of CP violation in top quark
  production and decay}},
  \href{https://doi.org/10.1016/0550-3213(92)90545-M}{\emph{Nucl. Phys. B}
  {\bfseries 388} (1992) 53}.

\bibitem{CMS:2022quh}
{\scshape CMS} collaboration, \emph{{Search for CP violating top quark
  couplings in pp collisions at $\sqrt{s}$ = 13 TeV}},
  \href{https://doi.org/10.1007/JHEP07(2023)023}{\emph{JHEP} {\bfseries 07}
  (2023) 023} [\href{https://arxiv.org/abs/2205.07434}{{\ttfamily
  2205.07434}}].

\bibitem{Garosi:2023yxg}
F.~Garosi, D.~Marzocca, A.R.~S{\'a}nchez and A.~Stanzione, \emph{{Indirect
  constraints on top quark operators from a global SMEFT analysis}},
  \href{https://doi.org/10.1007/JHEP12(2023)129}{\emph{JHEP} {\bfseries 12}
  (2023) 129} [\href{https://arxiv.org/abs/2310.00047}{{\ttfamily
  2310.00047}}].

\bibitem{Bisal:2025jwv}
S.~Bisal, \emph{{Constraining ALP-top interaction from the chromoelectric
  dipole moment of the top quark}},
  \href{https://doi.org/10.1103/w6wh-knnp}{\emph{Phys. Rev. D} {\bfseries 112}
  (2025) 055027} [\href{https://arxiv.org/abs/2507.12570}{{\ttfamily
  2507.12570}}].

\bibitem{Han:2024gan}
T.~Han, D.~Liu and S.~Wang, \emph{{Top quark electroweak dipole moment at a
  high energy muon collider}},
  \href{https://doi.org/10.1103/PhysRevD.111.035015}{\emph{Phys. Rev. D}
  {\bfseries 111} (2025) 035015}
  [\href{https://arxiv.org/abs/2410.11015}{{\ttfamily 2410.11015}}].

\bibitem{Janot:2015yza}
P.~Janot, \emph{{Top-quark electroweak couplings at the FCC-ee}},
  \href{https://doi.org/10.1007/JHEP04(2015)182}{\emph{JHEP} {\bfseries 04}
  (2015) 182} [\href{https://arxiv.org/abs/1503.01325}{{\ttfamily
  1503.01325}}].

\bibitem{Bernreuther:2017cyi}
W.~Bernreuther, L.~Chen, I.~Garc{\'\i}a, M.~Perell{\'o}, R.~Poeschl, F.~Richard
  et~al., \emph{{CP-violating top quark couplings at future linear $e^+e^-$
  colliders}}, \href{https://doi.org/10.1140/epjc/s10052-018-5625-3}{\emph{Eur.
  Phys. J. C} {\bfseries 78} (2018) 155}
  [\href{https://arxiv.org/abs/1710.06737}{{\ttfamily 1710.06737}}].

\bibitem{Subba:2026nzs}
A.~Subba and Y.~Shi, \emph{{Quantumness of top quark pairs produced at LHC
  within SMEFT framework}},  \href{https://arxiv.org/abs/2605.12033}{{\ttfamily
  2605.12033}}.

\bibitem{Alloul:2013bka}
A.~Alloul, N.D.~Christensen, C.~Degrande, C.~Duhr and B.~Fuks, \emph{{FeynRules
  2.0 - A complete toolbox for tree-level phenomenology}},
  \href{https://doi.org/10.1016/j.cpc.2014.04.012}{\emph{Comput. Phys. Commun.}
  {\bfseries 185} (2014) 2250}
  [\href{https://arxiv.org/abs/1310.1921}{{\ttfamily 1310.1921}}].

\bibitem{Degrande:2011ua}
C.~Degrande, C.~Duhr, B.~Fuks, D.~Grellscheid, O.~Mattelaer and T.~Reiter,
  \emph{{UFO - The Universal FeynRules Output}},
  \href{https://doi.org/10.1016/j.cpc.2012.01.022}{\emph{Comput. Phys. Commun.}
  {\bfseries 183} (2012) 1201}
  [\href{https://arxiv.org/abs/1108.2040}{{\ttfamily 1108.2040}}].

\bibitem{Alwall:2014hca}
J.~Alwall, R.~Frederix, S.~Frixione, V.~Hirschi, F.~Maltoni, O.~Mattelaer
  et~al., \emph{{The automated computation of tree-level and next-to-leading
  order differential cross sections, and their matching to parton shower
  simulations}}, \href{https://doi.org/10.1007/JHEP07(2014)079}{\emph{JHEP}
  {\bfseries 07} (2014) 079} [\href{https://arxiv.org/abs/1405.0301}{{\ttfamily
  1405.0301}}].

\bibitem{Ollivier:2001fdq}
H.~Ollivier and W.H.~Zurek, \emph{{Introducing Quantum Discord}},
  \href{https://doi.org/10.1103/PhysRevLett.88.017901}{\emph{Phys. Rev. Lett.}
  {\bfseries 88} (2001) 017901}
  [\href{https://arxiv.org/abs/quant-ph/0105072}{{\ttfamily
  quant-ph/0105072}}].

\bibitem{Henderson:2001wrr}
L.~Henderson and V.~Vedral, \emph{{Classical, quantum and total correlations}},
  \href{https://doi.org/10.1088/0305-4470/34/35/315}{\emph{J. Phys. A}
  {\bfseries 34} (2001) 6899}
  [\href{https://arxiv.org/abs/quant-ph/0105028}{{\ttfamily
  quant-ph/0105028}}].

\bibitem{Luo:2008ecu}
S.~Luo, \emph{{Quantum discord for two-qubit systems}},
  \href{https://doi.org/10.1103/PhysRevA.77.042303}{\emph{Phys. Rev. A}
  {\bfseries 77} (2008) 042303}.

\bibitem{Dakic:2010xfz}
B.~Daki{\'c}, V.~Vedral and v.~Brukner, \emph{{Necessary and Sufficient
  Condition for Nonzero Quantum Discord}},
  \href{https://doi.org/10.1103/PhysRevLett.105.190502}{\emph{Phys. Rev. Lett.}
  {\bfseries 105} (2010) 190502}
  [\href{https://arxiv.org/abs/1004.0190}{{\ttfamily 1004.0190}}].

\bibitem{Wiseman:2007hyt}
H.M.~Wiseman, S.J.~Jones and A.C.~Doherty, \emph{{Steering, Entanglement,
  Nonlocality, and the Einstein-Podolsky-Rosen Paradox}},
  \href{https://doi.org/10.1103/PhysRevLett.98.140402}{\emph{Phys. Rev. Lett.}
  {\bfseries 98} (2007) 140402}
  [\href{https://arxiv.org/abs/quant-ph/0612147}{{\ttfamily
  quant-ph/0612147}}].

\bibitem{Jevtic:2015epl}
S.~Jevtic, M.J.W.~Hall, M.R.~Anderson, M.~Zwierz and H.M.~Wiseman,
  \emph{{Einstein-Podolsky-Rosen steering and the steering ellipsoid}},
  \href{https://doi.org/10.1364/JOSAB.32.000A40}{\emph{J. Opt. Soc. Am. B}
  {\bfseries 32} (2015) A40} [\href{https://arxiv.org/abs/1411.1517}{{\ttfamily
  1411.1517}}].

\bibitem{Mahlon:2010gw}
G.~Mahlon and S.J.~Parke, \emph{{Spin Correlation Effects in Top Quark Pair
  Production at the LHC}},
  \href{https://doi.org/10.1103/PhysRevD.81.074024}{\emph{Phys. Rev. D}
  {\bfseries 81} (2010) 074024}
  [\href{https://arxiv.org/abs/1001.3422}{{\ttfamily 1001.3422}}].

\bibitem{Atwood:1991ka}
D.~Atwood and A.~Soni, \emph{{Analysis for magnetic moment and electric dipole
  moment form-factors of the top quark via $e^+ e^- \to t \bar t$}},
  \href{https://doi.org/10.1103/PhysRevD.45.2405}{\emph{Phys. Rev. D}
  {\bfseries 45} (1992) 2405}.

\bibitem{Diehl:1993br}
M.~Diehl and O.~Nachtmann, \emph{{Optimal observables for the measurement of
  three gauge boson couplings in $e^+ e^- \to W^+ W^-$}},
  \href{https://doi.org/10.1007/BF01555899}{\emph{Z. Phys. C} {\bfseries 62}
  (1994) 397}.

\bibitem{Bernreuther:1995nw}
W.~Bernreuther and P.~Overmann, \emph{{Probing Higgs boson and supersymmetry
  induced CP violation in top quark production by (un)polarized electron -
  positron collisions}}, \href{https://doi.org/10.1007/s002880050265}{\emph{Z.
  Phys. C} {\bfseries 72} (1996) 461}
  [\href{https://arxiv.org/abs/hep-ph/9511256}{{\ttfamily hep-ph/9511256}}].

\bibitem{Bartl:1998nn}
A.~Bartl, E.~Christova, T.~Gajdosik and W.~Majerotto, \emph{{CP violating
  energy asymmetries of b and anti-b quarks in $e^+ e^- \to t \bar t$}},
  \href{https://doi.org/10.1103/PhysRevD.59.077503}{\emph{Phys. Rev. D}
  {\bfseries 59} (1999) 077503}
  [\href{https://arxiv.org/abs/hep-ph/9803426}{{\ttfamily hep-ph/9803426}}].

\bibitem{Durieux:2018tev}
G.~Durieux, M.~Perell{\'o}, M.~Vos and C.~Zhang, \emph{{Global and optimal
  probes for the top-quark effective field theory at future lepton colliders}},
  \href{https://doi.org/10.1007/JHEP10(2018)168}{\emph{JHEP} {\bfseries 10}
  (2018) 168} [\href{https://arxiv.org/abs/1807.02121}{{\ttfamily
  1807.02121}}].

\bibitem{Sjostrand:2014zea}
T.~Sj{\"o}strand, S.~Ask, J.R.~Christiansen, R.~Corke, N.~Desai, P.~Ilten
  et~al., \emph{{An introduction to PYTHIA 8.2}},
  \href{https://doi.org/10.1016/j.cpc.2015.01.024}{\emph{Comput. Phys. Commun.}
  {\bfseries 191} (2015) 159}
  [\href{https://arxiv.org/abs/1410.3012}{{\ttfamily 1410.3012}}].

\bibitem{deFavereau:2013fsa}
{\scshape DELPHES 3} collaboration, \emph{{DELPHES 3, A modular framework for
  fast simulation of a generic collider experiment}},
  \href{https://doi.org/10.1007/JHEP02(2014)057}{\emph{JHEP} {\bfseries 02}
  (2014) 057} [\href{https://arxiv.org/abs/1307.6346}{{\ttfamily 1307.6346}}].

\bibitem{ILC:2013jhg}
{\scshape ILC} collaboration, H.~Baer et~al., eds., \emph{{The International
  Linear Collider Technical Design Report - Volume 2: Physics}},
  \href{https://arxiv.org/abs/1306.6352}{{\ttfamily 1306.6352}}.

\bibitem{ILCInternationalDevelopmentTeam:2022izu}
{\scshape ILC International Development Team} collaboration, \emph{{The
  International Linear Collider: Report to Snowmass 2021}},
  \href{https://arxiv.org/abs/2203.07622}{{\ttfamily 2203.07622}}.

\bibitem{CMS:2025dpp}
{\scshape CMS} collaboration, \emph{{Search for $CP$ violation in events with
  top quarks and Z bosons at $\sqrt{s}$ = 13 and 13.6 TeV}},
  \href{https://doi.org/10.1016/j.physletb.2025.139857}{\emph{Phys. Lett. B}
  {\bfseries 869} (2025) 139857}
  [\href{https://arxiv.org/abs/2505.21206}{{\ttfamily 2505.21206}}].

\bibitem{Hayreter:2015ryk}
A.~Hayreter and G.~Valencia, \emph{{T-odd correlations from the top-quark
  chromoelectric dipole moment in lepton plus jets top-pair events}},
  \href{https://doi.org/10.1103/PhysRevD.93.014020}{\emph{Phys. Rev. D}
  {\bfseries 93} (2016) 014020}
  [\href{https://arxiv.org/abs/1511.01464}{{\ttfamily 1511.01464}}].

\bibitem{Lamba:2026jfm}
P.~Lamba, F.~Maltoni, O.~Miniati and E.~Vryonidou, \emph{{Quantum detection of
  CP violation in the $t\bar{t}$ system: production}},
  \href{https://arxiv.org/abs/2607.25029}{{\ttfamily 2607.25029}}.

\bibitem{Lamba:2026sni}
P.~Lamba, F.~Maltoni, O.~Miniati and E.~Vryonidou, \emph{{Quantum detection of
  CP violation in the $t\bar{t}$ system: tomography}},
  \href{https://arxiv.org/abs/2607.25034}{{\ttfamily 2607.25034}}.

\end{thebibliography}\endgroup
\end{document}